\documentclass[aps,prb,twocolumn,superscriptaddress,floatfix]{revtex4-1}

\usepackage[percent]{overpic}
\usepackage{graphicx,graphics}
\usepackage{dcolumn}
\usepackage{textcomp} 
\usepackage{amsmath,amssymb,amsfonts}
\usepackage{latexsym,verbatim}
\usepackage{bm}
\usepackage{mathtools}
\usepackage{bbold}
\usepackage{color}
\usepackage{ulem}

\usepackage{braket}
\usepackage{soul}
\begin{document}
\title{Charger-mediated energy transfer for quantum batteries: \\ an open system approach}
\author{Donato Farina}
\email{donato.farina@sns.it}
\affiliation{Istituto Italiano di Tecnologia, Graphene Labs, Via Morego 30, I-16163 Genova, Italy}
\affiliation{NEST, Scuola Normale Superiore, I-56126 Pisa, Italy}
\author{Gian Marcello Andolina}
\affiliation{Istituto Italiano di Tecnologia, Graphene Labs, Via Morego 30, I-16163 Genova, Italy}
\affiliation{NEST, Scuola Normale Superiore, I-56126 Pisa, Italy}
\author{Andrea Mari}
\affiliation{NEST, Scuola Normale Superiore and Istituto Nanoscienze-CNR, I-56127 Pisa,~Italy}
\author{Marco Polini}
\affiliation{Istituto Italiano di Tecnologia, Graphene Labs, Via Morego 30, I-16163 Genova, Italy}
\author{Vittorio Giovannetti}
\affiliation{NEST, Scuola Normale Superiore and Istituto Nanoscienze-CNR, I-56127 Pisa,~Italy}

\date{\today}

\begin{abstract}
The energy charging of a quantum battery is analyzed in an open quantum setting, where the interaction between
the battery element and the external power source is mediated by an ancilla system (the quantum charger) that acts as a controllable switch. Different  implementations are analyzed putting emphasis on the interplay between coherent energy pumping mechanisms  and thermalization. 
\end{abstract}
\maketitle

\section{Introduction} 
A battery is a physical system that is capable to store energy supplied by an external source, making it available to other devices. Its performance is characterized by several figures of merit gauging the amount of energy it can store and/or deliver as a function of its mass/volume and how these quantities vary over time. Motivated by the constant progress of miniaturization of electronic devices and stimulated by the success obtained in other sectors by adopting 
analogous approaches~\cite{Riedel2017,Acin17}, increasing interest has been recently devoted to analyze the performances of
``quantum batteries", i.e.~energy storing systems which, at least in principle, could exploit genuine quantum effects to obtain improved performances 
with respect to conventional (say classical) schemes~\cite{Alicki13, Hovhannisyan13, Binder15, Campaioli17,Ferraro17,Le17,ALTRO,ALTROERGO}.

The core of this idea ultimately relies on the 
 possibility of achieving superior performances in the manipulation of energy by cleverly exploiting
 quantum resources~\cite{Vinjanampathy16,Alicki18,Goold16,Campisi11,Horodecki13,Gelbwaser-Klimovsky15,Strasberg16}. 
 Starting from the seminal, but abstract works of Refs.~\onlinecite{Alicki13, Hovhannisyan13,Binder15, Campaioli17}, concrete implementations  of  quantum batteries  have been proposed~\cite{Ferraro17,Le17}.
 At the same time, more sophisticated  modelizations of the charging process have been presented~\cite{ALTRO,ALTROERGO} which put emphasis  on the problems that could arise at the interface between a quantum battery B and its 
 external energy supply A, the ``quantum charger" (also modelled as a quantum system). In particular, in Ref.~\onlinecite{ALTROERGO} 
it was pointed out that quantum correlations between  B and A, while possibly playing an important role in speeding up
the charging of the battery,  could  result in a net detrimental effect by reducing the amount of energy 
that one could transform in useful work once having access to B alone 
(a reasonable scenario in any relevant practical applications). 
Building up from these observations, in the present work we introduce a further generalization 
of the quantum battery/quantum charger model by explicitly embedding the whole system into an external environment E whose action is effectively described in terms of a master equation~\cite{BreuerPetruccione2007}.
\begin{figure}[t]
\centering
\begin{overpic}[width=0.8\columnwidth]{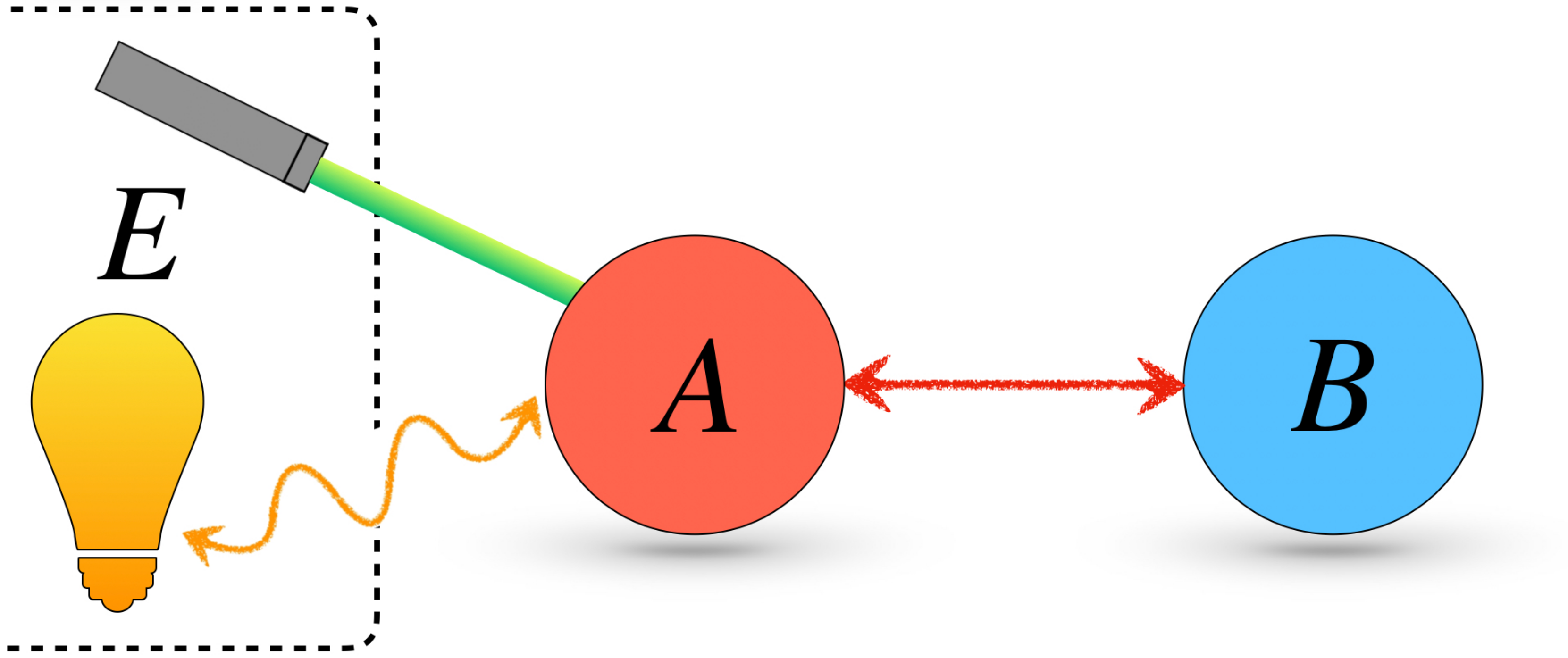}
\end{overpic}
\caption{(Color online) Pictorial representation of the model analyzed in this work. 
Here, energy from the external world E  flows into the ancillary system A, which acts as a classical-to-quantum
transducer for B (the quantum battery). 
The subsystems A and B interact via a time-dependent
coupling, which is switched on during the charging interval $[0,\tau]$ only.
In our model, the EA coupling may either occur via the interaction with a thermal source  (represented by the yellow lamp), or  coherently
via the modulation of the local Hamiltonian of A  (represented by the green laser), or both.
  \label{fig:open-scheme}}
\end{figure}
Accordingly, and at variance with previous proposals~\cite{Ferraro17,ALTRO,ALTROERGO}, in our
approach 
the energy meant to be transferred to the quantum battery is not assumed to sit initially 
on the charger A. Instead, it is dynamically injected into the system thanks to the presence of E, either via
thermalization or via coherent driving induced by external control, the ancilla A merely playing the role of an effective transducer capable to convert such inherently classical inputs into ``quantum signals'' for B.
In this context,  for different implementations of the A and B systems, 
we explicitly compute the total energy transferred to the battery and, using the results of Refs.~\onlinecite{ERGO,ALTROERGO},
the fraction of it that turns out to be useful in terms of extractable work.
Specifically, we are interested in studying the different ways in which the thermal and coherent driving mechanisms contribute to the 
process, enlightening possible cross-talking effects between the two. 
Interestingly enough, while typically the presence of thermal pumping tends to reduce the fraction of stored energy which
can be extracted as work, in some implementations which exhibit effective nonlinearities in the coupling between A and B, we find evidences of a positive interplay which, for an assigned
intensity of the coherent driving force, tends to increase the performances of the quantum battery, an effect which is reminding us of the  noise assisted energy transfer observed in quantum biology~\cite{BIO1,BIO2}.

Our Article is organized as following. In Sect.~\ref{TheoryOpen} we introduce the general model and the
figures of merit we are going to analyze. Sect.~\ref{subsec:hoho-open} reports the results we obtained when both the charger  A and the battery B are harmonic oscillators, while Sect.~\ref{sect:open-TLS} deals instead with the
two-qubit scenario. Finally, results for the hybrid case where A is a harmonic oscillator and B is a qubit are reported in Sect.~\ref{SEC:TRE}. A brief summary and our main conclusions are reported in Sect.~\ref{Conclusions}. Useful technical details can be found in Appendices~\ref{ERGOGO}-\ref{appendix-QubQub-me}.

\section{General theory} 
\label{TheoryOpen}

The model we are interested in studying  consists in three separate elements:
a quantum battery B, an external energy supply E, and an ancillary quantum system  A that acts as mediator between the other two elements, 
see Fig.~\ref{fig:open-scheme}. Alternatively, one can interpret A as  that part of a structured global bath AE, which is directly interacting with B, E representing instead the nonlocal degree of freedoms of the environment. 
In our treatment we shall represent A and B as actual quantum systems  whose dynamics is determined by a  Markovian Master Equation (ME) which effectively 
accounts for the  presence of E.  
We thus describe the temporal evolution of the density matrix  $\rho_{\rm A B}(t) $ of the AB system as ($\hbar=1$ throughout this Article):
\begin{eqnarray}\label{eq:me-qub-qubSCHEMA}
\dot{\rho}_{\rm A B}(t) &=& - i \left[{H}_{\rm A} +{H}_{\rm B}
,\rho_{\rm AB}(t)\right]  + \lambda(t) {\cal L}_{\rm AB}(t)[\rho_{\rm A B}(t) ]~, \nonumber \\
\end{eqnarray}
where $[\cdots, \cdots]$ denotes the usual commutator.
In Eq.~(\ref{eq:me-qub-qubSCHEMA}), the first term contains the  free Hamiltonian 
of the system composed by the local (time-independent) contributions of A and B which,
for sake of convenience, 
we shall assume to have zero ground-state energy. 
The second term, instead, is explicitly time dependent and refers to the AB interactions and to the  charging terms of the model induced by the coupling between the external energy supply E and A.
Here $\lambda(t)$ is a   dimensionless 
function  equal to $1$ for $t\in[0,\tau[$ and $0$ elsewhere, which we use for turning  ``on/off" such contributions, $\tau$ representing
the charging time of the protocol. ${\cal L}_{\rm AB}(t)$ is instead a  Gorini-Kossakolski-Sudarshan-Lindblad (GKSL) super-operator~\cite{Kossakowski72,Lindblad76,Gorini76} that contains both coherent and dissipative contributions.
Explicitly, we write it as
\begin{eqnarray}
{\cal L}_{\rm AB}(t)[ \cdots ] \equiv 
 - i \left[\Delta{H}_{\rm A}(t)+{H}^{(1)}_{\rm AB} ,\cdots \right]  +  {\cal D}^{(T)}_{\rm A}[\cdots]~, \nonumber \\ \label{LIND}
\end{eqnarray}
where $H^{(1)}_{{\rm AB}}$ is the interaction Hamiltonian between the charger and the battery,
$\Delta{H}_{\rm A}(t)$ is a local modulation of the energy of A which is externally driven by classical fields that may inject energy into the system, and finally ${\cal D}^{(T)}_{\rm A}$ is a purely dissipative contribution that acts locally on A and which accounts 
for the local thermalization  of A induced by a bosonic bath at temperature $T$ (no direct dissipation being assumed for B). 
In this scenario we assume that for $t<0$, when A and B do not interact and are isolated from the rest, they are prepared in the ground state of the local terms  
$H_{\rm A}$ and $H_{\rm B}$ respectively, i.e.~
\begin{eqnarray} \label{eq:in-cond-qubqub}
\rho_{\rm AB}(t \le 0)=\ket{0}\Bra{0}_{\rm A}\otimes \ket{0}\Bra{0}_{\rm B}~.
\end{eqnarray} 
At time $t=0$, A is attached to the external supply E  by switching on the dissipator ${\cal D}_{\rm A}^{(T)}$ and  (possibly)  the modulation   $\Delta{H}_{{\rm A}}(t)$, while A and B begin to interact with each other.
In the time window $[0, \tau[$ part of the energy coming from the outside---and going only to A at short time scales---flows to B thanks to the non-zero internal  coupling term ${H}^{(1)}_{{\rm AB}}$, which we assume to commute with the free Hamiltonian $H_{\rm A}+H_{\rm B}$,
\begin{eqnarray} \label{COMM} 
[ {H}^{(1)}_{{\rm AB}}, H_{\rm A}+H_{\rm B}]=0~.
\end{eqnarray}
At the end of the charging process, namely at time $\tau$ when $\lambda(t)$ returns to zero, we isolate again the system and turn the interaction between A and B off.

\subsection{Figures of merit}
\label{SEC:FIGs}

Under the above conditions, we are interested in characterizing how efficiently energy can be transferred into the battery.
For this purpose, we study the mean energy contained in B at the end of the charging process and the corresponding ergotropy~\cite{ERGO}, i.e.,  respectively,  the quantities
\begin{eqnarray}\label{stored energy}
 E_{\rm B}(\tau)&\equiv&{\rm tr}[{H}_{\rm B} \rho_{\rm B}(\tau)]~,  \\ 
 \mathcal{E}_{\rm B}(\tau) &\equiv& E_{\rm B}(\tau)  - \min_{\substack{U}_{\rm B}}{\rm tr}\left[ {H}_{\rm B} U_{\rm B} \rho_{\rm B}(\tau)  U_{\rm B}^\dagger \right]~,\label{eq:ergo}
\end{eqnarray}
where $\rho_{\rm B}(\tau) \equiv {\rm tr}_{\rm A} [\rho_{\rm AB}(\tau)]$ is the reduced state of the battery
at time $\tau$, and where minimization in Eq.~(\ref{eq:ergo}) is performed over all the unitaries $U_{\rm B}$ acting locally on such system. 
The first of these functions measures 
 the total amount of energy that has been transferred to B thanks to the mediation of
the charger A. The second, instead, provides us with the part 
of $E_{\rm B}(\tau)$ which can be turned into work 
 while having access to the battery alone, a reasonable scenario in many  applications where  A and  E are not available to a generic end user~\cite{ALTROERGO}. Indeed, it may happen that part of the mean energy  of B 
will be locked into correlations between such system and the charging device, 
 preventing one from accessing it via local operations on the battery. 
The term  we are subtracting from $E_{\rm B}(\tau)$ in right-hand-side of Eq.~(\ref{eq:ergo}) 
exactly targets such contributions. It formally corresponds to
 the expectation value 
$E^{(p)}_{\rm B}(\tau) \equiv {\rm tr}[{H}_{\rm B} \rho^{(p)}_{\rm B}(\tau)]$ 
 of ${H}_{\rm B}$ computed on the 
passive state~\cite{ERGO,PASSIVE} 
$\rho^{(p)}_{\rm B}(\tau)$,  obtained by properly reordering the spectrum of  $\rho_{\rm B}(\tau)$ and replacing the associated eigenvectors with  those of the system Hamiltonian---see Appendix~\ref{ERGOGO} for details.

In what follows, we shall analyze the quantities $E_{\rm B}(\tau)$ and ${\mathcal E}_{\rm B}(\tau)$, 
their ratio
\begin{eqnarray}\label{ratio} 
R_{\rm B}(\tau) \equiv {\mathcal E}_{\rm B}(\tau)/E_{\rm B}(\tau)~, 
\end{eqnarray}
as well as their
associated mean charging  powers
\begin{eqnarray} 
P_{\rm B}(\tau)&\equiv &  E_{\rm B}(\tau)/\tau\label{POWER}~, \\ 
\mathcal{P}_{\rm B}(\tau)&\equiv &  \mathcal{E}_{\rm B}(\tau)/\tau\label{POWERERGO}~, 
\end{eqnarray}
for different choices of A and B systems and for different  energy-injection mechanisms. 
For all these models we shall enforce resonant conditions of the local energies of A and B, as well as for the driving term $\Delta H_{\rm A}(t)$. 
 This will allow us to simplify the analysis by solving the ME in the time interval $[0,\tau]$ in the interaction picture representation where instead
 of $\rho_{\rm AB}(t)$ one focuses on its rotated version
 \begin{eqnarray}
 \tilde{\rho}_{\rm AB}(t) \equiv e^{i (H_{\rm A}+H_{\rm B}) t}
{\rho}_{\rm AB}(t)  e^{-i (H_{\rm A}+H_{\rm B}) t}~, \label{INTPICT}
\end{eqnarray}
for which Eq.~(\ref{eq:me-qub-qubSCHEMA}) for $t\in[0, \tau[$  reduces to 
 \begin{eqnarray}
\dot{\tilde{\rho}}_{\rm A B}(t)&=&{\cal L}_{\rm AB}[\tilde{\rho}_{\rm A B}(t) ]~.
\label{eq:me-qub-qubSCHEMAfree}
\end{eqnarray}
Here, ${\cal L}_{\rm AB}$ is as in (\ref{LIND}) but with $\Delta{H}_{\rm A}(t)$ replaced by the constant term $\Delta{H}_{\rm A}\equiv \Delta{H}_{\rm A}(t=0)$. 

Most importantly, under the above conditions, 
  both the mean energy 
(\ref{stored energy}) and the ergotropy  (\ref{eq:ergo}) of B will be then directly computed on 
the reduced density matrix $\tilde{\rho}_{\rm B}(\tau) =  {\rm tr}_{\rm A} [\tilde{\rho}_{\rm AB}(\tau)]$ of 
$\tilde{\rho}_{\rm AB}(\tau)$. Indeed, the latter  differs from 
 ${\rho}_{\rm B}(\tau)$ by a unitary rotation induced by $H_{\rm B}$, i.e. 
$\tilde{\rho}_{\rm B}(\tau) = e^{i H_{\rm B} \tau}
{\rho}_{\rm B}(\tau)  e^{-i H_{\rm B} \tau}$. Accordingly, we have 
${\rm tr}[{H}_{\rm B} \tilde{\rho}_{\rm B}(\tau)] = {\rm tr}[{H}_{\rm B} {\rho}_{\rm B}(\tau)] = E_{\rm B}(\tau)$
while, including $e^{i H_{\rm B} \tau}$ into the minimization over $U_{\rm B}$, we have
 \begin{eqnarray} 
\min_{\substack{U}_{\rm B}}{\rm tr}\left[ {H}_{\rm B} U_{\rm B} \tilde{\rho}_{\rm B}(\tau)  U_{\rm B}^\dagger \right] =
\min_{\substack{U}_{\rm B}}{\rm tr}\left[ {H}_{\rm B} U_{\rm B} {\rho}_{\rm B}(\tau)  U_{\rm B}^\dagger \right], \nonumber 
\end{eqnarray}
which, via Eq.~(\ref{eq:ergo}), ensures that $\tilde{\rho}_{\rm B}(\tau)$ and ${\rho}_{\rm B}(\tau)$ possess
the same ergotropy value.

\section{Two-harmonic oscillator model}
\label{subsec:hoho-open}

We  begin by considering  the case in which both the charger A and the  quantum battery B are 
described by resonant harmonic oscillators assuming the following definitions for the Hamiltonian contribution to Eq.~(\ref{eq:me-qub-qubSCHEMA})
\begin{eqnarray}\label{eq:hoho-H-open}
&{H}_{\rm A}= \omega_0 a^\dagger a~, \qquad 
{H}_{\rm B} = \omega_0 b^\dagger b~,& \nonumber\\
&\Delta{H}_{\rm A}(t)= F \left( e^{-i \omega_0 t } a^\dagger + e^{i \omega_0 t } a \right)~, &\nonumber  \\ 
&{H}_{\rm AB}^{(1)}= g
\left( a b^\dagger + a^\dagger b\right)~.&
\end{eqnarray}
Here, $a$, $b$ (resp. $a^\dag$, $b^\dag$) are the bosonic annihilation (creation) operators of the A and B systems, respectively, $\omega_0$ is the fundamental frequency of the model, while 
$g$ and $F$ are
coupling constants gauging the AB coupling and the driving field acting on A. 
Regarding the GKSL dissipator we take
\begin{eqnarray}
{\cal D}_{\rm A}^{(T)} \equiv \gamma (N_{b}(T)+1) \mathcal{D}_{\rm A}^{[a]}+  \gamma N_{b}(T)\mathcal{D}_{\rm A}^{[a^\dag]}~,
\label{eq:me-qub-qub111}
\end{eqnarray}
where the rate $\gamma$ fixes the time scale of the dissipation process,
\begin{equation}\label{BoseEinstein}
N_{b}(T)\equiv \frac{1}{\exp{[\omega_{0}/(k_{\rm B} T)]}-1}
\end{equation}
is the mean number of bath quanta at frequency $\omega_0$,
and, given a generic operator $x_{\rm A}$ acting on A, $\mathcal{D}_{\rm A}^{[x]}$ represents the super-operator~\cite{BreuerPetruccione2007}
\begin{equation}\label{dissipators}
\mathcal{D}_{\rm A}^{[x]} [\cdots]\equiv  
x_{\rm A}  \cdots x_{\rm A}^\dagger -\frac{1}{2}\left \{ x_{\rm A}^\dagger x_{\rm A} , \cdots \right \}~, 
\end{equation}
with $\{\cdots , \cdots \}$ the anti-commutator symbol.
With this choice, the first term on the r.h.s. of Eq.~(\ref{eq:me-qub-qub111})
describes energy flow from the system into the environment with spontaneous and stimulated emission terms, whereas the second one describes energy flow from the environment into the system and, consistently,  gives a finite contribution to the ME only at non-zero temperatures. 

The associated interaction-picture-representation ME~(\ref{eq:me-qub-qubSCHEMAfree}) in this case reads
\begin{eqnarray}
&&\dot{\tilde{\rho}}_{\rm A B}(t)=-i \left[ g
\left( a b^\dagger + a^\dagger b\right)+F  ( a^\dagger + a ),\; \tilde{\rho}_{\rm A B}(t) \right]  \label{eq:ho-ho-lindblad-me} \\ 
&&\;+ \gamma(N_{b}(T)+1)  \mathcal{D}_{\rm A}^{[a]}\left[ \tilde{\rho}_{\rm A B}(t)  \right]
+ \gamma N_{b}(T) \mathcal{D}_{\rm A}^{[ a^\dag]} \left[\tilde{\rho}_{\rm A B}(t)  \right]~,\nonumber 
\end{eqnarray}
which admits explicit integration. 
In particular, since the generator on the r.h.s. of Eq.~(\ref{eq:ho-ho-lindblad-me}) is quadratic in the field modes, the dynamics preserves the Gaussian character~\cite{GAUSSIANCV} of the ground state~(\ref{eq:in-cond-qubqub}), which in this case is the zero Fock state of the A and B modes. Accordingly, a complete characterization of  $\tilde{\rho}_{\rm A B}(t)$ can be obtained by simply determining the first and second momenta of the field operators, i.e.~by solving the sets (\ref{eq:hoho-lang}), (\ref{eq:hoho-lang1}), and~(\ref{hoho-lang-3}) of coupled linear differential equations.
Specifically, 
 using $\left\langle x\right\rangle \equiv \mbox{tr}[ x \tilde{\rho}_{\rm AB}(t)]$ to indicate the expectation value of  a generic operator $x$ on $\tilde{\rho}_{\rm AB}(t)$, for the first momenta we have
\begin{equation}\label{eq:hoho-lang}\begin{cases}
\dot{\left\langle a\right\rangle}=-i (g\langle b\rangle+F) - {\displaystyle \frac{\gamma}{2}} \langle a\rangle~, \\
\dot{\langle b \rangle}=- i g \langle a\rangle~,
\end{cases}
\end{equation}
while, for the second momenta, we have
\begin{equation}\label{eq:hoho-lang1}\begin{cases}
 \dot{\langle a b^\dagger \rangle}=i \left[g( \langle a^\dagger a \rangle -  \langle b^\dagger b\rangle)- F \langle b \rangle^* \right]- {\displaystyle \frac{\gamma}{2}} \langle a b^\dagger\rangle~,\\
\dot{\langle b^\dagger b \rangle}=2g~{\rm Im}{\langle a b^\dagger\rangle}~,\\
\dot{\langle a^\dagger a  \rangle}= -2~{\rm Im}[g \langle a b^\dagger\rangle + F\langle a\rangle]- \gamma \langle a^\dagger a\rangle + \gamma N_{b}(T)~, 
   \end{cases}
\end{equation}
and
\begin{equation}
\begin{cases}
\dot{\left< a^2\right>}=-2 i (g \left< ab\right> + F \left< a\right>)-\gamma \left< a^2\right>~,\\
\dot{\left< ab\right>}=-i[g (\left< a^2\right>+\left< b^2\right>)+F\left<b\right>]- {\displaystyle\frac{\gamma}{2}}\left< ab\right>~,\\
\dot{\left< b^2\right>}=-2ig \left< ab\right>~. \label{hoho-lang-3}
\end{cases}
\end{equation}
The above differential equations, together with the initial conditions associated with (\ref{eq:in-cond-qubqub}), i.e.
\begin{eqnarray}\label{eq:hoho-lang-in-cond}
\langle a^\dagger a  \rangle\big\rvert_{t=0} &=&
 \langle b^\dagger b  \rangle\big\rvert_{t=0}=
\langle a^2  \rangle\big\rvert_{t=0}=
 \langle b^2  \rangle\big\rvert_{t=0}=0 \nonumber\\
\langle a \rangle\big\rvert_{t=0} &=&
 \langle b \rangle\Big\rvert_{t=0} =
  \langle a b^\dagger \rangle\big\rvert_{t=0}=
  \langle a b \rangle\big\rvert_{t=0}=0~,
\end{eqnarray} 
 are all we need to solve for the  evaluation of the figures of merit introduced in Sec.~\ref{SEC:FIGs}.
 In particular, $E_{\rm B}(\tau)$ simply corresponds to $\omega_0 \langle b^\dag b\rangle|_{t=\tau}$
 while for the ergotropy we can use the fact that $\tilde{\rho}_{\rm B}(\tau)$ is  Gaussian 
 so that we can use the results of Appendix~\ref{ERGOGO} to express it as 
\begin{equation}
\mathcal{E}_{\rm B}(\tau)=\omega_0 \left.\left(\left<b^\dagger b \right>-\frac{\sqrt{D}-1}{2}
\right)\right|_{t=\tau}~,
\label{eq:ergo-hoho}
\end{equation}
with 
\begin{equation}\label{DEFdet}
D\equiv \left(1+ 2 \left<b^\dagger b\right>  - 2 \left|\left<b\right> \right|^2\right)^2- 4 \left| \left<b^2\right> -\left<b\right> ^2\right|^2~.
\end{equation}
\subsection{Analysis} \label{Sec.ANA1}

The model  exhibits an effective decoupling between thermal and coherent pumping, which is reflected by the fact that,
for assigned values of $T$ and $F$,  
each of the functions $\langle x\rangle$ entering the Eqs.~(\ref{eq:hoho-lang})--(\ref{hoho-lang-3}) 
 can be expressed as the sum of two contributions,
\begin{eqnarray} \label{DEC1} 
\langle x\rangle = \left.\langle x\rangle\right\rvert_{F=0,T} + \left\langle x\rangle\right\rvert_{F,T=0}~,
\end{eqnarray} 
with $\langle x\rangle\rvert_{F=0,T}$ describing the solution of the differential equations in the absence of the coherent driving terms (i.e. with $F=0$),
and with  $\langle x\rangle\rvert_{F,T=0}$ describing instead the solution of the same equations with a thermal bath at
zero temperature (i.e. $T=0$)---see Appendix~\ref{DECOUPLING}. As a consequence of (\ref{DEC1}), for generic values of $T$ and $F$  we have
\begin{equation} \label{DECEB} 
\left.E_{\rm B}(\tau)\right\rvert_{F, T} =\left.E_{\rm B}(\tau)\right\rvert_{F=0, T} + \left.E_{\rm B}(\tau)\right\rvert_{F, T=0}~.
\end{equation}
An analogue simplification can also be observed for the ergotropy ${\mathcal{E}}_{\rm B}(\tau)$.
Indeed, notwithstanding the
fact that such quantity  has a nonlinear dependence on the first and second momenta of the fields operators---see Eqs.~(\ref{eq:ergo-hoho}) and (\ref{DEFdet})---only the contribution associated with the coherent
driving at zero temperature matters, i.e.
 \begin{equation}\label{DECERGOCOE}
\left.\mathcal{E}_{\rm B}(\tau)\right\rvert_{F, T} = \left.\mathcal{E}_{\rm B}(\tau)\right\rvert_{F, T=0} = 
\left.{E}_{\rm B}(\tau)\right\rvert_{F, T=0}~, \quad \mbox{$\forall$ $T\geq 0$~,}
\end{equation}
the ergotropy of the purely thermal driving case being always null, i.e.
 \begin{equation}\label{DECERGOTHER}
\left.\mathcal{E}_{\rm B}(\tau)\right\rvert_{F=0, T}=0~, \qquad \mbox{$\forall$ $T\geq 0$~.}
\end{equation}
(Explicit proofs of the above expressions, as well as  the derivation of  the same  relations which can be established for the local energy and ergotropy of the charger A, can be found in  Appendix~\ref{DECOUPLING}.) 

\begin{figure}
\centering
\begin{overpic}[width=0.8\columnwidth]{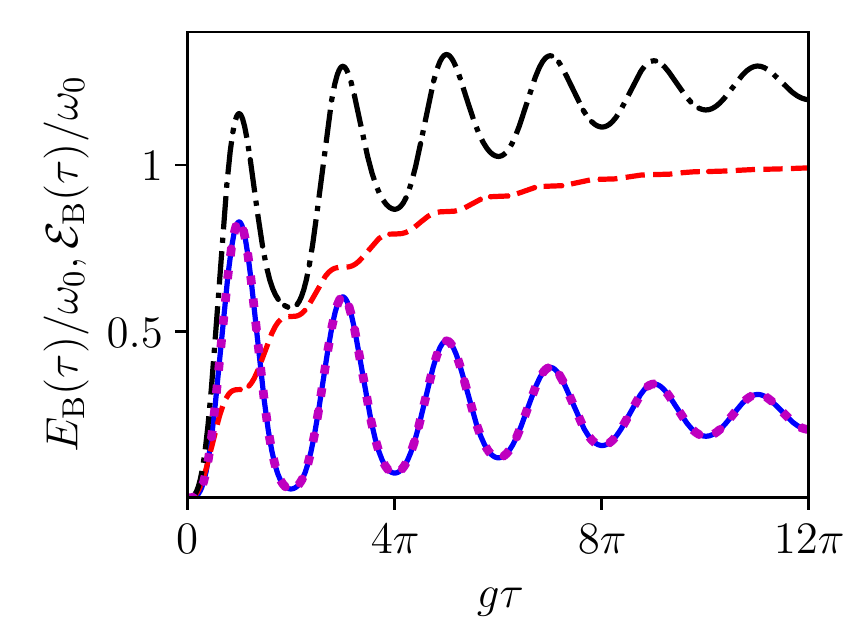}\put(0,75){\normalsize (a)}\end{overpic}\vspace{0.5em}
\begin{overpic}[width=0.8\columnwidth]{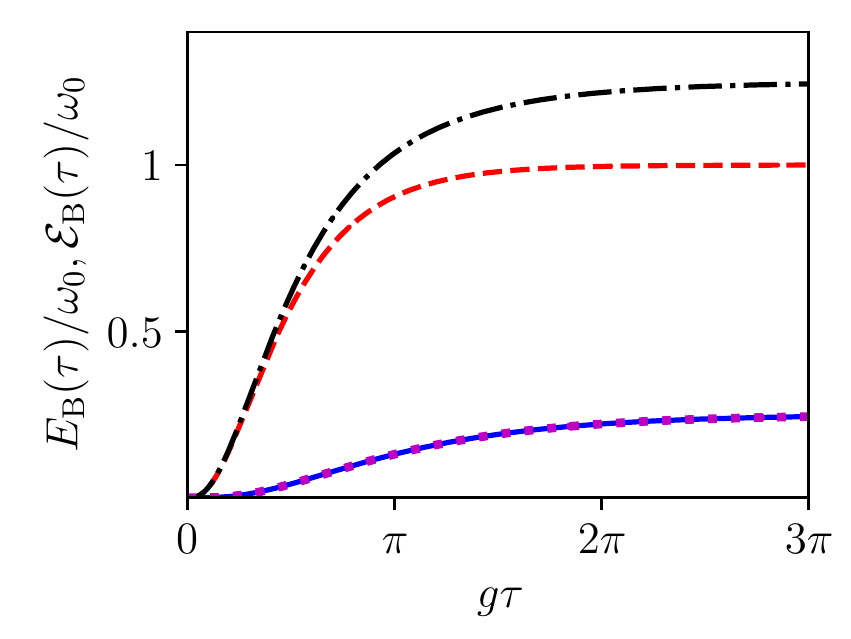}\put(0,75){\normalsize (b)}\end{overpic}\vspace{0.5em}
\caption{(Color online) Local energy ${E}_{\rm B}(\tau)$ and ergotropy $\mathcal{E}_{\rm B}(\tau)$ of the battery B (both in units of $\omega_{0}$) as functions of $g\tau$, 
for the two-harmonic oscillator model.  Panel (a) The black dash-dotted, red dashed, and magenta dotted curves represent $E_{\rm B}(\tau)$ for $N_{b}(T)=1$ and $F=0.1 \omega_0$, 
$N_b(T)=1$ and $F=0$ (no coherent driving), and $N_b(T)=0$ and $F=0.1 \omega_0$ (no thermal driving), respectively.
The blue solid curve represents the ergotropy $\mathcal{E}_{\rm B}(\tau)$ 
for $N_b(T)=1$ and $F=0.1 \omega_0$. Note that this curve is superimposed to the magenta one: this is because,  
as emphasized in Eq.~(\ref{DECERGOCOE}), $\left.\mathcal{E}_{\rm B}(\tau)\right\rvert_{F, T} = 
\left.{E}_{\rm B}(\tau)\right\rvert_{F, T=0}$. All numerical results in panel (a) have been obtained by setting $g=0.2 \omega_0$ and 
$\gamma = 0.05\omega_0$ (underdamped regime). Panel (b) Same as in panel (a) but for $\gamma = \omega_{0}$ (overdamped regime).
 \label{FIGharmosci}}
\end{figure}

\begin{figure}
\begin{overpic}[width=.8\columnwidth]{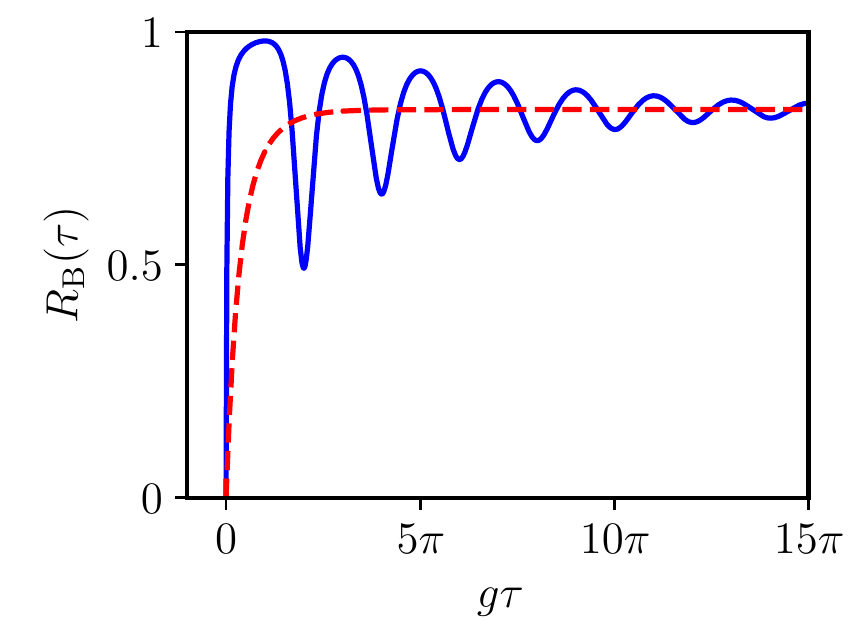}\put(0,75){(a)}\end{overpic}\vspace{0.5em}
\begin{overpic}[width=.8\columnwidth]{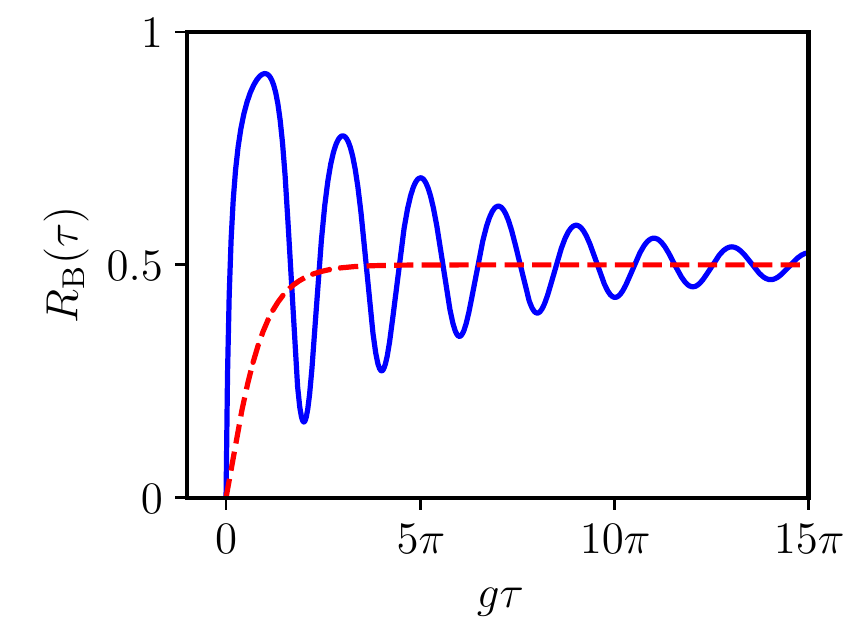}\put(0,75){(b)}\end{overpic}\vspace{0.5em}
\caption{(Color online) The ratio~(\ref{ratio}), which measures the fraction of energy stored in the battery which can be extracted as work, as a function of $g\tau$ and for the two-harmonic oscillator model. Panel (a) Different curves correspond to different values of the loss parameter $\gamma$. Blue solid line: $\gamma = 0.05\omega_0$ (underdamped regime); $\gamma =\omega_0$ (overdamped regime): red dashed line. The other parameters have been set as following: $F=0.2 \omega_0$, $g=0.2 \omega_0$, $N_b(T)=0.2$.
Panel (b) Same as in panel (a) but for $N_b(T)=1$. $R_{\rm B}$ decreases for increasing temperature. In both panels, all curves approach the asymptotic value (\ref{ASYMP111}) for $\tau \gg 1/g$. \label{FIGratio}}
\end{figure}

Equations~(\ref{DECEB})--(\ref{DECERGOTHER}) 
represent an important simplification, which allows us to address the functional dependence
upon $T$ and $F$ of ${E}_{\rm B}(\tau)$ and  $\mathcal{E}_{\rm B}(\tau)$ 
by studying separately their effects on the battery model.
 This is a peculiarity of the two-harmonic-oscillator model, which is not found 
 in different implementations  where instead  one witnesses a non-trivial interplay between the coherent and thermal driving contributions---see next sections. 
In the present case, the above identities 
imply that  while non-zero values of  $T$ and $F$  both add 
 to ${E}_{\rm B}(\tau)$, only the $F$ matters in the transferring of energy that is 
 useful for future  extractions of work. (A non-zero bath temperature can 
only decrease the ratio~(\ref{ratio})
 but cannot deteriorate the net value of the ergotropy  associated with a given choice of~$F$.)
 Anticipating the analytic solutions we present in the  coming subsections,
   examples of these behaviours can be found in  Figs.~\ref{FIGharmosci} and~\ref{FIGratio}---the first displaying the functional dependence of 
 ${E}_{\rm B}(\tau)$ and $\mathcal{E}_{\rm B}(\tau)$ upon $\tau$ for various combinations of $N_b(T)$ and $F$, while
 the second presenting instead the ratio $R_{\rm B}(\tau)$ for two different  bath temperatures---and in the 
 asymptotic values attained by ${E}_{\rm B}(\tau)$, $\mathcal{E}_{\rm B}(\tau)$ 
in the $\tau\rightarrow \infty$ limit, i.e.
\begin{eqnarray}
{E}_{\rm B}(\infty)&=& \omega_0  N_b(T) + \omega_0 (F/g)^2~, \nonumber \\ 
 \mathcal{E}_{\rm B}(\infty)&=& \omega_0 (F/g)^2~,
\end{eqnarray}
whose associated ratio~(\ref{ratio})
\begin{eqnarray}
{R}_{\rm B}(\infty)&=& \frac{F^2}{ g^2 N_b(T) 
+F^2}~,\label{ASYMP111} 
 \end{eqnarray}
clearly exhibits a monotonic decreasing behaviour with respect to $T$.

\subsubsection{Thermal energy supply regime ($F=0$, $T$ generic)}  
\begin{figure*}
\centering
\begin{tabular}{ccc}
\begin{overpic}[width=0.67\columnwidth]{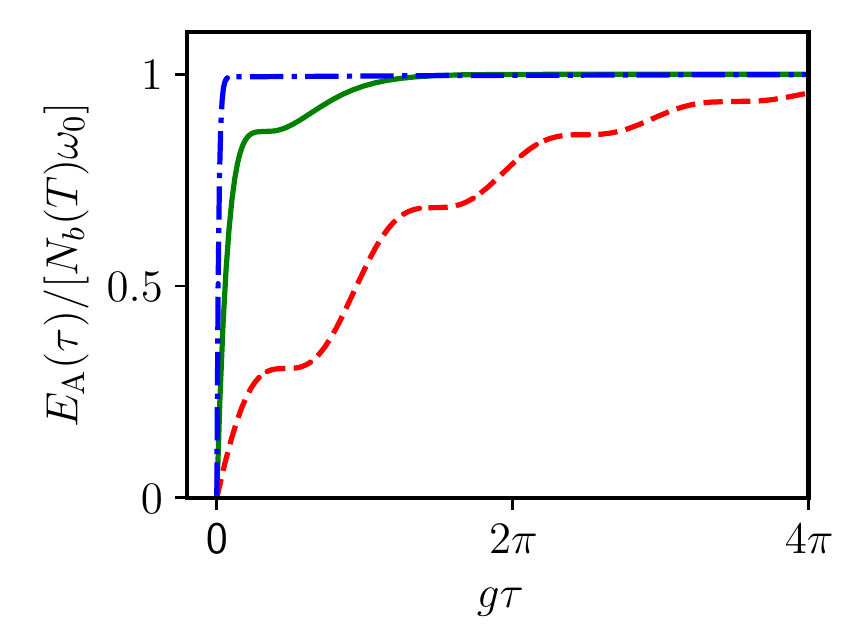}\put(10,75){\normalsize (a)}\end{overpic}\vspace{0.5em} & 
\begin{overpic}[width=0.67\columnwidth]{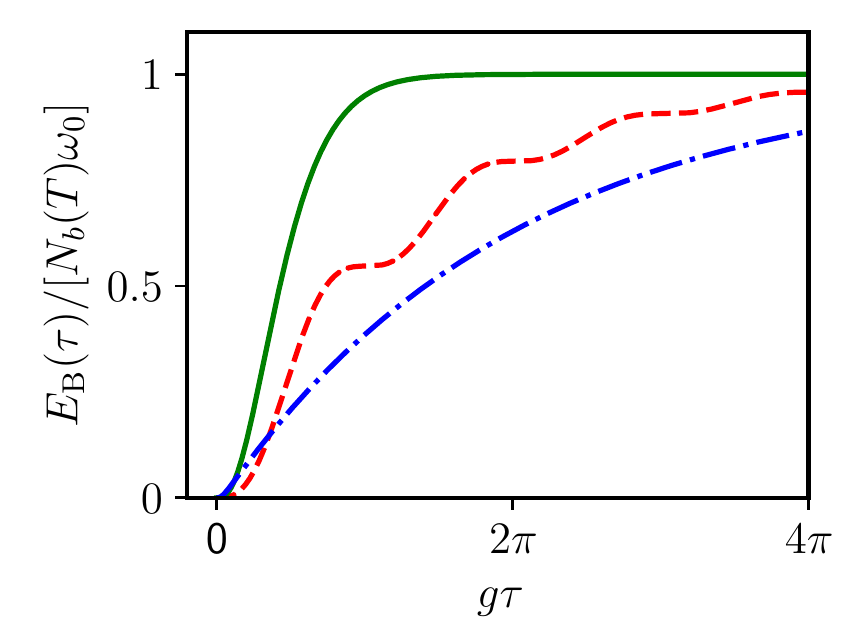}\put(10,75){\normalsize (b)}\end{overpic}\vspace{0.5em} &
\begin{overpic}[width=0.67\columnwidth]{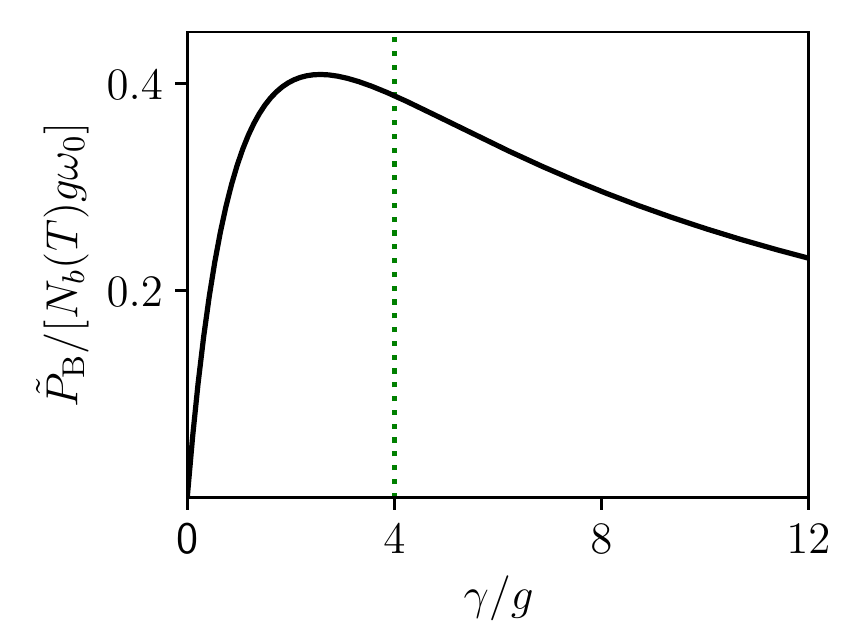}\put(10,75){\normalsize (c)}\end{overpic}\vspace{0.5em} 
\end{tabular}
\caption{(Color online) Panel (a) The local energy ${E}_{\rm A}(\tau)$ of the ancilla A (in units of $N_{b}(T)\omega_{0}$)---Eq.~(\ref{eq:hoho-EaEb_I})---as a function of $g\tau$ for the two-harmonic oscillator model. Different curves correspond to different values of the ratio $\gamma/g$.
Red dashed line: $\gamma/g=1/2$; green solid line: $\gamma/g=4$; blue dash-dotted line: $\gamma/g=25$. Panel (b) Same as in panel (a) but for the energy $E_{\rm B}(\tau)$ stored in the battery B---Eq.~(\ref{eq:hoho-EaEb_II}). Panel (c) The maximum average storing power $\tilde{P}_{\rm B}$ (in units of $g \omega_0N_b(T)$)---Eq.~(\ref{MAXPO})---as a function of $\gamma/g$. All results in (a)-(c) have been obtained for the purely energy supply regime (i.e.~$F=0$).
 \label{FIGnew3}}
\end{figure*}

Let us consider first the case where no coherent driving is present (i.e. $F=0$) while
A is in contact with a non-zero temperature bath (i.e. $T>0$).  
As anticipated in Eq.~(\ref{DECERGOCOE}), this regime represents a poor implementation of the charging of a quantum battery 
 as it results in a zero value for the  ergotropy $\mathcal{E}_{\rm B}(\tau)$.
 For what concerns the mean energy of B, 
 setting
\begin{eqnarray}\label{DEFEP} 
\epsilon \equiv \sqrt{\gamma^2-(4g)^2}~,
\end{eqnarray}
explicit integration of 
Eqs.~(\ref{eq:hoho-lang})-(\ref{eq:hoho-lang-in-cond}) yields
\begin{eqnarray}\label{eq:hoho-EaEb_II}
\left.E_{\rm B}(\tau)\right\rvert_{F=0,T}& =& \omega_0  N_{b}(T)  \Big\{ 1+ (e^{-\frac{1}{2}\gamma \tau} /\epsilon^2 ) \\
\nonumber   
&\times&
\left[16 g^2  - \gamma \epsilon \sinh(\epsilon \tau /2) - \gamma^2 \cosh(\epsilon \tau /2) \right]\Big\}~,
\end{eqnarray}
to be compared with 
\begin{eqnarray}\label{eq:hoho-EaEb_I}
\left.E_{\rm A}(\tau)\right\rvert_{F=0,T}& =&\omega_0  N_{b}(T)  \Big\{ 1+ (e^{-\frac{1}{2}\gamma \tau} /\epsilon^2 )\\ \nonumber   
&\times&
\left[ 16 g^2  + \gamma \epsilon \sinh(\epsilon \tau /2) - \gamma^2 \cosh(\epsilon \tau /2) \right]\Big\}~,
\end{eqnarray}
which instead represents the mean local energy of the ancillary system A at the end of the process.
In the limit of large $\tau$, Eqs.~(\ref{eq:hoho-EaEb_II}) and (\ref{eq:hoho-EaEb_I}) show convergency of 
$E_{\rm A}(\tau)$ and $E_{\rm B}(\tau)$
toward the same value $\omega_0  N_{b}(T)$, in agreement with the (local) thermalization of the two subsystems. 
The transient, however, exhibits two distinct regimes: an oscillating underdamped regime occurring for $\gamma <4g$, and an overdamped regime for $\gamma\geq 4g$ characterized by a monotonic increment of $E_{\rm B}(\tau)$, which for large enough $\gamma$ 
can be conveniently approximated as  $E_{\rm B}(\tau)\approx \omega_0 N_{b}(T)  (1-e^{-4g^2\tau/\gamma})$,
see panels a) and b)  of Fig.~\ref{FIGnew3}. 
This feature has a profound  impact on the  timing  of the process: 
a numerical analysis reveals in fact that  the charging time of the battery (defined e.g. as the first time at which B reaches a given fraction of its asymptotic value $\omega_0 N_b(T)$), exhibits a non trivial dependence upon the parameters  $\gamma$ and $g$ with optimal performances
attained when they are close to the critical point $\gamma =4g$. 
A clear evidence of this phenomenon can be found by looking at the maximum  of the average storing power (\ref{POWER}),
\begin{eqnarray} \label{MAXPO}
\tilde{P}_{\rm B} \equiv  \max_\tau P_{\rm B}(\tau)~,
\end{eqnarray}
which, as shown in  panel c) of Fig.~\ref{FIGnew3},  acquires its largest value just below threshold. 
We anticipate that the same effect will be observed in all the other implementations we discuss in the remaining of this Article, at least when 
the coherent driving is not present (i.e. $F=0$). 
 A possible  explanation of  the arising of such fine tuning condition between $\gamma$ and $g$ in the optimization of the 
 charging process 
can  be found by noticing that while 
   the battery needs a finite loss coefficient to be thermally excited, a too large value of the loss coefficient will tend to freeze the state of A via an environment-mediated  quantum Zeno effect~\cite{BreuerPetruccione2007}, 
   preventing the latter to  efficiently transfer energy to B. 

 \subsubsection{Coherent energy supply regime ($T=0$, $F$ generic)}

 Consider next the scenario where $F\neq 0$ and the bath temperature is zero, i.e.  $T=0$. 
  From Eqs.~(\ref{DECEB}) and (\ref{DECERGOCOE}), it follows that this is the optimal setting in terms  of our ability
  of maximizing the fraction of energy stored in B, which is available for work extraction at later times. Indeed, in this case we have 
\begin{eqnarray} \label{IMPO111} 
\left.\mathcal{E}_{\rm B}(\tau)\right\rvert_{F, T=0} = \left.{E}_{\rm B}(\tau)\right\rvert_{F, T=0} \label{FFD}~, \end{eqnarray}
  corresponding to the optimal value $1$ for the ratio~(\ref{ratio})---the same identity applying also for the energy that resides on A, i.e. $\mathcal{E}_{\rm A}(\tau)\rvert_{F, T=0}={E}_{\rm A}(\tau)\rvert_{F, T=0}$, see Eq.~(\ref{DECEA}).
 This result is a consequence of the fact
 that   in the $T=0$ regime 
  the system  AB remains in a factorized, pure coherent state at all times, see Eq.~(\ref{FACT1}) of the Appendix. Specifically, we have
\begin{eqnarray} 
\tilde{\rho}_{\rm AB}(\tau) = \ket{\alpha(\tau)}_{\rm A}\bra{\alpha(\tau)} \otimes \ket{\beta(\tau)}_{\rm B}\bra{\beta(\tau)}\;, \label{FACT10} 
 \end{eqnarray}
where,  given $\epsilon$ as in Eq.~(\ref{DEFEP}), $\alpha(\tau)$ and $\beta(\tau)$ are the following 
coherent amplitudes
\begin{eqnarray}\label{eq:hoho-coh-I11} 
\alpha(\tau)&=&   -i\frac{4  F}{\epsilon} e^{-\frac{\gamma \tau}{4}} \sinh({\epsilon\tau }/{ 4})~, \\ \nonumber 
\beta(\tau)&=&   -\frac{F}{g}  \left\{1-e^{-\frac{\gamma \tau }{ 4}}
\left[\cosh( {\epsilon \tau }/{ 4})+ \frac{\gamma}{\epsilon}\sinh({\epsilon \tau }/{ 4})\right]
\right\}~.
\end{eqnarray}
The associated local mean energies are hence given by 
\begin{eqnarray}\label{eq:hoho-coh-I}
&&\left.E_{\rm A}(\tau)\right\rvert_{F,T=0}=\omega_0 |\alpha(\tau)|^2\\  \nonumber 
&&\qquad\qquad\qquad
=  \frac{16 ~\omega_0 F^2}{\epsilon^2} e^{-\frac{\gamma\tau}{2}} \sinh ^2({\epsilon \tau }/{ 4})
\end{eqnarray}
and
\begin{eqnarray} \label{eq:hoho-coh-II} 
&&\left.E_{\rm B}(\tau)\right\rvert_{F,T=0} =\omega_0 |\beta(\tau)|^2 \\ \nonumber 
&&\quad =\frac{ \omega_0F^2}{g^2}  \left\{1-e^{-\frac{\gamma \tau }{ 4}}
\left[\cosh( {\epsilon \tau }/{ 4})+ \frac{\gamma}{\epsilon}\sinh({\epsilon \tau }/{ 4})\right]
\right\}^2~, 
\end{eqnarray}
which, thanks to Eqs.~(\ref{FFD}) and (\ref{DECEA}), coincide  with the  ergotropies $\mathcal{E}_{\rm A}(\tau)$ and $\mathcal{E}_{\rm B}(\tau)$ of the two systems. 
One may observe that, for all non-zero values of the damping parameter $\gamma$,    in the limit  $\tau\rightarrow \infty$ the energy of A nullifies testifying that the ancilla asymptotically approaches its local ground-state eigenstate, while the coherent amplitude of B  reaches
a finite  value $\beta(\infty)= - F/g$. 
As this result is non-perturbative in $g$, the energy stored in B in this regime can become very large resulting in
\begin{eqnarray} \label{HARMASYMP} 
E_{\rm B}(\infty)\Big\rvert_{F,T=0} = { \omega_0}(F/g)^2\;,\end{eqnarray} with the charger A going back to the initial vacuum state after a transient. (Notice that is 
formally equivalent to directly attaching the driving to the battery B and putting it in contact with a zero-temperature thermal bath, with loss coefficient $2g$ rather than $\gamma$.)
 
\begin{figure}
\centering
\begin{overpic}[width=0.8\columnwidth]{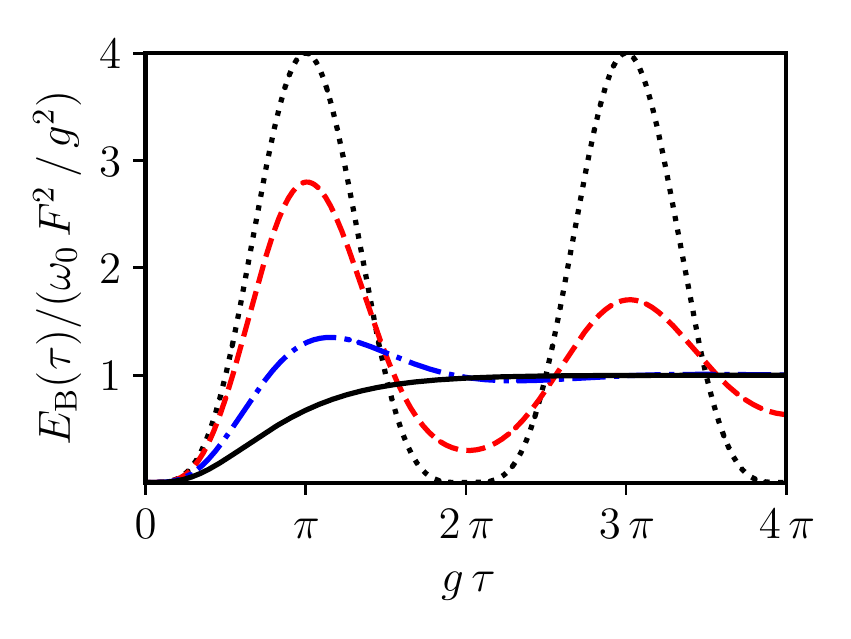}\put(0,75){(a)}\end{overpic}\vspace{0.5em}
\begin{overpic}[width=0.8\columnwidth]{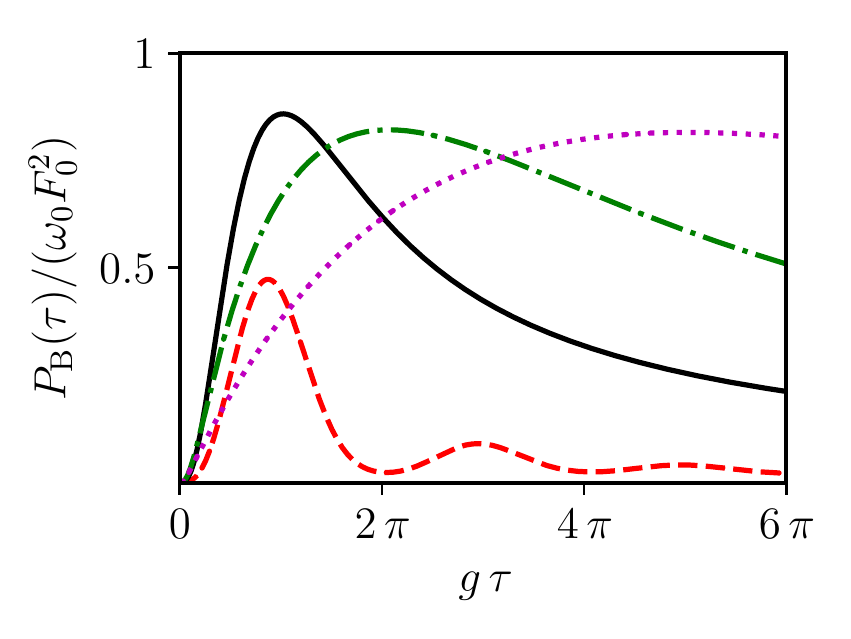}\put(0,75){(b)}\end{overpic}\vspace{0.5em}
\caption{(Color online) Panel (a) $E_{\rm B}(\tau)$ (in units of $\omega_{0} F^2/g^2$)---Eq.~(\ref{eq:hoho-coh-II})---as a function of $g\tau$ for the two-harmonic oscillator model. Different curves correspond to different values of the loss parameter $\gamma$. Black dotted line: $\gamma=0$; red dashed line: $\gamma=0.1\omega_0$; 
blue dash-dotted line: $\gamma=0.4\omega_0$; black solid line: $\gamma=0.8\omega_0$. Results have been obtained by setting $g=0.2\omega_0$. Panel (b) Same as in panel (a) but for $F=F_{0}\sqrt{\gamma}$. Different curves correspond to different values of the loss parameter $\gamma$. Red dashed line: $\gamma=0.1\omega_0$; black solid line: $\gamma=0.8\omega_0$; green dash-dotted line: $\gamma=2\omega_0$; magenta dotted line: $\gamma=5\omega_0$. We notice that in this case  $\gamma$ needs to be tuned with $g$ in order to get high power in a short time. All results in this figure refer to the case of the purely coherent energy supply regime, i.e.~$N_{b}(T)=0$.
\label{fig:hoho-PsGamma}}
\end{figure}

The way this asymptotic configuration is attained  is not influenced by the specific  value of $F$ which in Eqs.~(\ref{eq:hoho-coh-I11}), (\ref{eq:hoho-coh-I}) 
appears as a multiplicative factor and does not affect the time scales---see Fig.~\ref{fig:hoho-PsGamma}. As discussed in Ref.~\onlinecite{ALTRO}, this peculiarity stems from the nature of the spectrum at hand, which is not upper bounded.
What instead plays an important role in  the transient is once more the ratio between $\gamma$ and $g$
which, 
as in the purely thermal energy supply scenario we analyzed before, can again be used to identify underdamped 
($\gamma <4g$) and overdamped ($\gamma\geq 4g$) regimes. Furthermore, as evident from 
 panel (a) of Fig.~\ref{fig:hoho-PsGamma}, it is clear that losses tend to reduce the value of the maximum energy. The best configuration is approached for $\gamma\rightarrow 0$ where the energy dynamics of B becomes periodic in $\tau$, i.e. 
$E_{\rm B}(\tau)=4 \omega_0 F^2\sin^4(g\tau/2)/g^2$,
allowing the battery to reach an energy (and ergotropy) level which can be up to  four times larger than the asymptotic value $E_{\rm B}(\infty)$, the smallest driving time $\tau$  ensuring this result being $\pi/g$. 
Under the same condition, a numerical evaluation shows that the associated energy storing power (\ref{POWER})  exhibits a maximum value 
$\tilde{P}_{\rm B}$  equal to $0.33 \times \left(4\,\omega_0\,F^2/g\right)$ for
an optimal charging time 
$\sim 2.78/g$. This is rather different from what we witnessed in the purely thermal setting where instead
the largest possible value of $\tilde{P}_{\rm B}$ was attained for values of $\gamma$ close to the threshold point---see panel c) of Fig.~\ref{FIGnew3}.  A possible reconciliation of this discrepancy can be found by noting that 
in realistic models  the quantities $F$ and $\gamma$ cannot be treated as independent parameters. 
For instance considering a standard cavity-QED implementation of the model,  from the microscopic derivation of the Lindblad equation~\cite{BreuerPetruccione2007}, it is more correct to assume $F\simeq F_0 \sqrt{\gamma}$, indicating that the more the laser is able to pump energy in the system the more the system would be subject to losses, being it more strongly coupled with the external world.
In such case, the analogy with the purely thermal setting is restored as it turns out that one must tune $\gamma$ with $g$ to obtain the highest charging power ${P}_{\rm B}(\tau)$ in the shortest time---see  panel b) of Fig.~\ref{fig:hoho-PsGamma}.

\section{Two-qubit model}
\label{sect:open-TLS}

In this Section we consider  the case in which both the charger A and the  quantum battery B are (resonant) two-level systems.
For this purpose, indicating with $\omega_0$ the level spacing of A and B,  we set the Hamiltonian terms entering  the ME~(\ref{eq:me-qub-qubSCHEMA}) as 
\begin{eqnarray}\label{eq:qubqub-H-open}
&{H}_{\rm A}= {\displaystyle \frac{\omega_0}{2}} \left(\sigma_{\rm{A}}^z+1\right)~, \qquad 
{H}_{\rm B} = {\displaystyle \frac{\omega_0}{2}} \left(\sigma_{\rm{B}}^z+1\right)~,& \nonumber\\
&\Delta{H}_{\rm A}(t)= F ( e^{-i\omega_0 t}\,\sigma^+_{\rm{A}} + e^{i\omega_0 t} \sigma^-_{\rm{A}} )~, &\nonumber  \\ 
&{H}_{\rm AB}^{(1)}=  g
\left(\sigma^-_{\rm{A}} \sigma^+_{\rm{B}} + \sigma^+_{\rm{A}} \sigma^-_{\rm{B}}\right)~,&
\end{eqnarray}
where for ${\rm X}={\rm A,B}$, $\sigma_{\rm X}^{x,y,z}$ represent the Pauli matrices acting on the system ${\rm X}$, and where 
$\sigma^+_{\rm{X}} = [\sigma^-_{\rm{X}}]^\dag= (\sigma^x_{\rm{X}}+i \sigma^y_{\rm{X}})/2$ is the corresponding two-level raising operator. 

In the above expressions, $\Delta{H}_{\rm A}(t)$ describes an external resonant driving field 
with amplitude $F$ while  ${H}_{\rm AB}^{(1)}$ is an exchange coupling term characterized by the coupling constant $g$ that commutes with the free component of the Hamiltonian. 
Regarding the GKSL dissipator  ${\cal D}_{\rm A}^{(T)}$ of (\ref{LIND}),
setting $N_b(T)$ as in Eq.~(\ref{BoseEinstein}) and $\mathcal{D}_{\rm A}^{[x]}$ as in Eq.~(\ref{eq:me-qub-qub111}) we take
\begin{eqnarray}
{\cal D}_{\rm A}^{(T)} \equiv \gamma (N_{b}(T)+1) \mathcal{D}_{\rm A}^{[\sigma^-]}+ \gamma N_{b}(T)\mathcal{D}_{\rm A}^{[\sigma^+]}~,
 \label{eq:me-qub-qub}
\end{eqnarray}
$\gamma$ fixing once again the time scale of the dissipation process.

In the interaction picture representation, the corresponding  ME~(\ref{eq:me-qub-qubSCHEMAfree}) of the problem explicitly reads as 
\begin{eqnarray}
&&\dot{\tilde{\rho}}_{\rm A B}(t)=-i \left[g \left(\sigma^-_{\rm{A}} \sigma^+_{\rm{B}} + \sigma^+_{\rm{A}} \sigma^-_{\rm{B}} \right)+F  (\sigma^+_{\rm{A}} +    \sigma^-_{\rm{A}}),\; \tilde{\rho}_{\rm A B}(t) \right]\nonumber \\
&&\;+ \gamma(N_{b}(T)+1)  \mathcal{D}_{\rm A}^{[\sigma^-]}\left[ \tilde{\rho}_{\rm A B}(t)  \right]
+ \gamma N_{b}(T) \mathcal{D}_{\rm A}^{[ \sigma^+]} \left[\tilde{\rho}_{\rm A B}(t)  \right]~,\nonumber \\ \label{eq:at-at-lindblad-me}
\end{eqnarray}
which admits  analytical integration, e.g.  
representing the operators in a given basis and obtaining a Cauchy problem for a system of linear ordinary differential equations.
Technical details are reported in Appendix~\ref{appendix-QubQub-me} while here we comment the obtained solutions for the stored energy $E_{\rm B}(\tau)$ and the associated ergotropy $\mathcal{E}_{\rm B}(\tau)$. 
Regarding these quantities,  it is worth observing that, for the choice~(\ref{eq:qubqub-H-open}) of the local Hamiltonian $H_{\rm B}$,
Eqs.~(\ref{stored energy}) and (\ref{eq:ergo}) yield
\begin{eqnarray}
E_{\rm B}(\tau) &=&\frac{\omega_0}{2}\left[1+r_z(\tau)\right] \label{stored energy1}~, \\ 
\mathcal{E}_{\rm B}(\tau)&=&\frac{\omega_0}{2}\left[{r}(\tau)+r_z(\tau)\right]~,\label{eq:ergo1} 
\end{eqnarray}
where ${r}(\tau) \equiv |\vec{r}(\tau)|$ and $r_z(\tau)$ represent, respectively, the length and the $z$-component of the
Bloch vector $\vec{r}(\tau)$ associated with the density matrix  $\tilde{\rho}_{\rm B}(\tau)$ of B---see Eq.~(\ref{eq:ergo1IMPO}) of  Appendix~\ref{ERGOGO} for a derivation of this result.

\subsection{Analysis} 
\label{sec.ANA2} 

\begin{figure*}
\centering
\begin{tabular}{cc}
\begin{overpic}[width=.8\columnwidth]{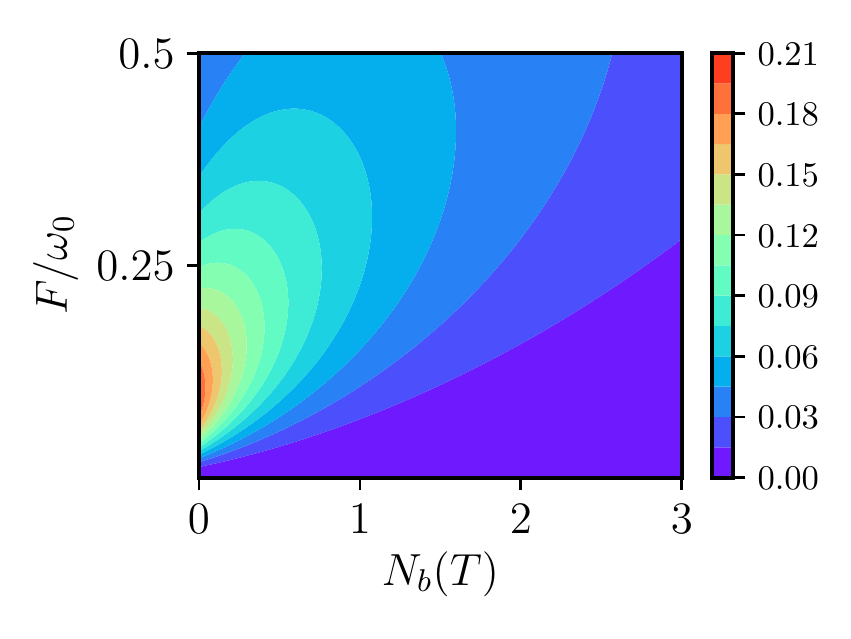}\put(10,75){\normalsize (a)}\end{overpic}\vspace{0.5em} & 
\begin{overpic}[width=.8\columnwidth]{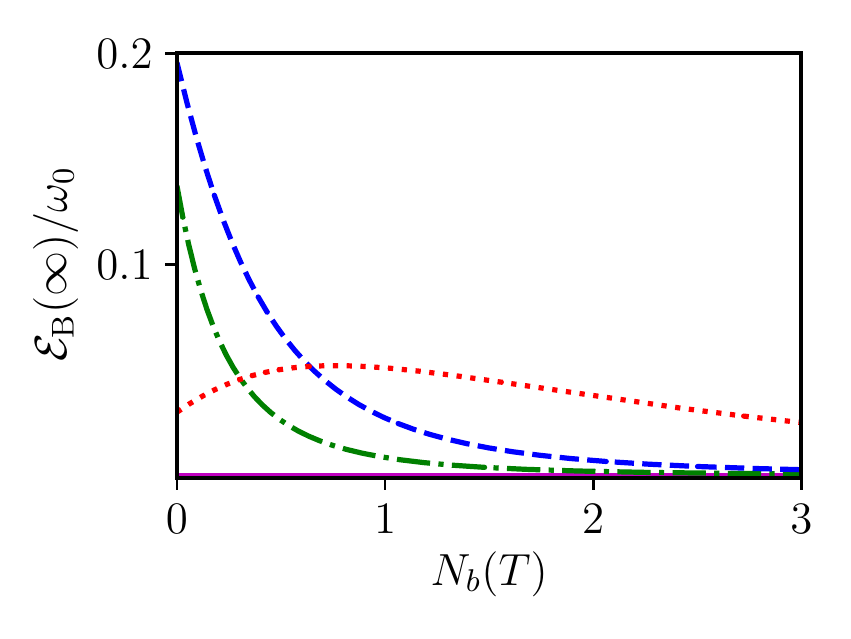}\put(10,75){\normalsize (b)}\end{overpic}\vspace{0.5em} \\
\begin{overpic}[width=.8\columnwidth]{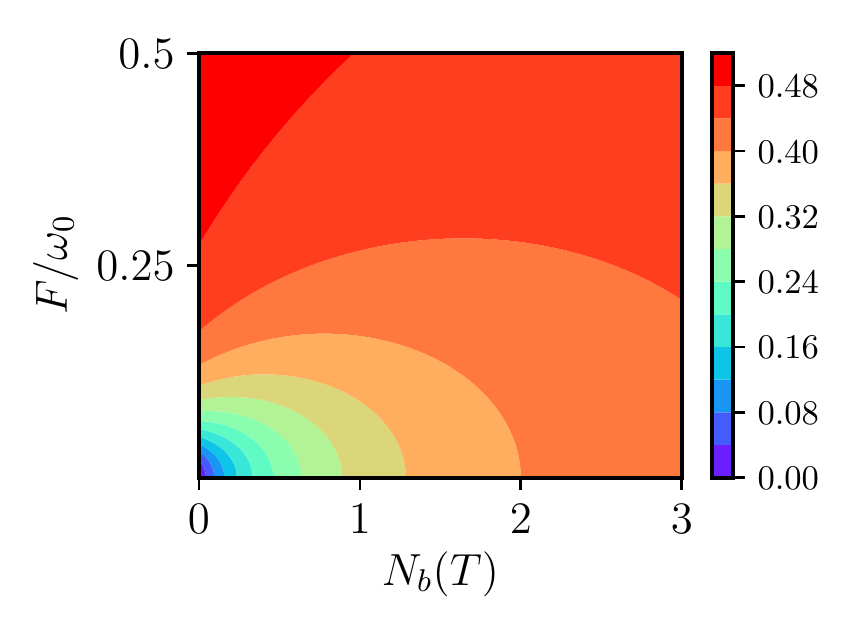}\put(10,75){\normalsize (c)}\end{overpic}\vspace{0.5em} & 
\begin{overpic}[width=.8\columnwidth]{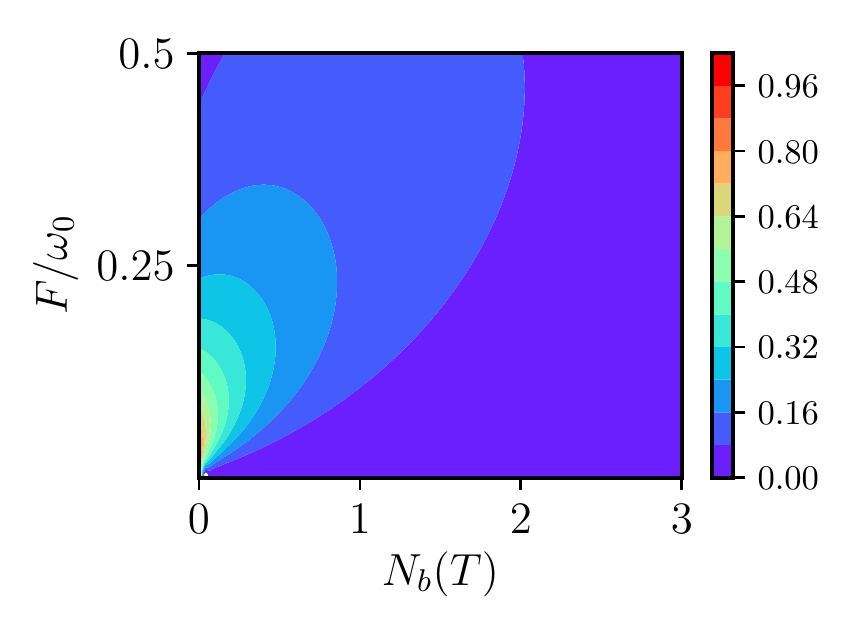}\put(10,75){\normalsize (d)}\end{overpic}\vspace{0.5em}
\end{tabular}
\caption{(Color online) Panel (a) Two-dimensional color plot of ${\mathcal E}_{\rm B}(\infty)$ (in units of $\omega_{0}$)---Eq.~(\ref{ERGOASYMP1})---as a function of $F$ (in units of $\omega_{0}$) and $N_{b}(T)$ for the two-qubit model. Notice that $\mathcal{E}_{\rm B}(\infty)$  reaches its
maximum value (\ref{MAXE}) for $N_b(T)=0$ (zero temperature) and $F\simeq 1.09 g$. For large enough $F$ we notice that
$\mathcal{E}_{\rm B}(\infty)$ is not monotonically decreasing in $N_b(T)$. Panel (b) ${\mathcal E}_{\rm B}(\infty)$ (in units of $\omega_{0}$) as a function of $N_{b}(T)$. 
Different curves correspond to different values of $F$. Magenta solid line: $F = 0$ (which yields ${\mathcal E}_{\rm B}(\infty)=0$); green dash-dotted line: $F =0.05\omega_{0}$; blue dashed line: $F=0.1\omega_{0}$; red dotted line: $F= 0.5\omega_{0}$. The non-monotonic behaviour as a function of 
$N_{b}(T)$ is clearly evident for $F=0.5\omega_{0}$.
Panel (c) Same as in panel (a) but for the asymptotic value $E_{\rm B}(\infty)$ of the energy stored in B---Eq.~(\ref{EASYMP1}).
 Panels d) Same as in panels (a) and (c) but for the ratio $R_{\rm B}(\infty)$---Eq.~(\ref{ratio}) in the $\tau \to \infty$ limit. $R_{\rm B}(\infty)$ reaches its maximum value for $N_{b}(T)=0$ and in the $F\to 0$ limit.
 All results in this figure have been obtained by setting $g=0.1\omega_0 $ and $\gamma=\omega_0$. \label{PLOTERGOASYMPMIX}}
\end{figure*}

Since at low energy the 
two-harmonic oscillator model discussed in the previous Section
has similar spectral properties to those of the two-qubit setting, we expect 
 the two schemes to exhibit  analogous performances 
in the low supply limit, i.e.~for $F\ll g,\gamma$ and $k_{\rm B}T\ll \omega_0$.
On the contrary, for not negligible values of $F$ or $T$, 
the effective nonlinearities introduced by the finite dimensionality of
the two-qubit model we are considering here, result in 
a more complex interplay between the coherent and incoherent pumping mechanisms  than the one we discussed
in Sect.~\ref{subsec:hoho-open}. 
Specifically, as will shall see, while still 
one cannot achieve non-zero values of $\mathcal{E}_{\rm B}(\tau)$ in the absence of the external coherent driving (i.e. $F=0$),
 decoupling rules similar to the ones reported in Eqs.~(\ref{DECERGOCOE}) and (\ref{DECERGOTHER}) hold no longer for arbitrary values of the system parameters. 
In particular, it turns out that, at variance with the two-harmonic oscillator model, the presence of a non-zero temperature
can strongly interfere with the ergotropy production. 
 Interestingly enough, while typically such interference tends to
 reduce $\mathcal{E}_{\rm B}(\tau)$, there are special settings of the system parameters for which one observes that a non-zero temperature 
can indeed result in a larger value of the attainable ergotropy. 
Evidences  of such behaviours can be obtained by looking at the values that  $E_{\rm B}(\tau)$ and $\mathcal{E}_{\rm B}(\tau)$ attain in the asymptotic $\tau\rightarrow \infty$ limit, which can be extrapolated from 
Eq.~(\ref{eq:at-at-lindblad-me}) by enforcing the stationary condition $\dot{\tilde{\rho}}_{\rm A B}(t)=0$.
The resulting expressions for arbitrary values of $T$ and $F$ in this case are given by
\begin{widetext}
\begin{equation}
\frac{E_{\rm B}(\infty)}{\omega_0}=\frac{1}{2}-\frac{g^2 \gamma \Gamma  (2 g^2 + \Gamma^2)}{32 F^4 (2 g^2 + \gamma^2)+4 F^2 \gamma ^2
   \left[  24 N_b(T) (N_b(T)+1) g^2 +(2 g^2 + \Gamma^2)   \right]+2 g^2 \Gamma ^2  (2 g^2 + \Gamma^2)}\,,
   \label{EASYMP1} 
\end{equation}
\begin{equation}
\frac{\mathcal{E}_{\rm B}(\infty)}{\omega_0} =
\frac{g \gamma  (2 g^2 + \Gamma^2)(\sqrt{4\gamma^2 F^2+ g^2 \Gamma^2 }-g \Gamma )
   }{32 F^4 (2 g^2 + \gamma^2)+4 F^2 \gamma ^2 \left[  24 N_b(T) (N_b(T)+1) g^2 + (2 g^2 + \Gamma^2)   \right]+2 g^2 \Gamma ^2  (2 g^2 + \Gamma^2)}\,, \label{ERGOASYMP1} 
\end{equation}
\end{widetext}
where $\Gamma$ is the renomalization of the loss coefficient $\gamma$ by the Bose occupation number 
$N_b(T)$ of the bath, i.e.
\begin{eqnarray}\label{DEFGAMMA} 
{\Gamma}\equiv \gamma (2 N_{b}(T) + 1)~.
\end{eqnarray}
In Fig.~\ref{PLOTERGOASYMPMIX} we display the functional dependence of the functions~(\ref{EASYMP1}) and (\ref{ERGOASYMP1}) 
  and of their ratio $R_{\rm B}(\infty)= \mathcal{E}_{\rm B}(\infty)/{E}_{\rm B}(\infty)$
in terms of $N_b(T)$ and $F$. As evident from panels a) and b) of the figure,
when $F$ is sufficiently large,
$\mathcal{E}_{\rm B}(\infty)$ may indeed take advantage  from an increase of the bath temperature.

\begin{figure}
\begin{overpic}[width=0.8\columnwidth]{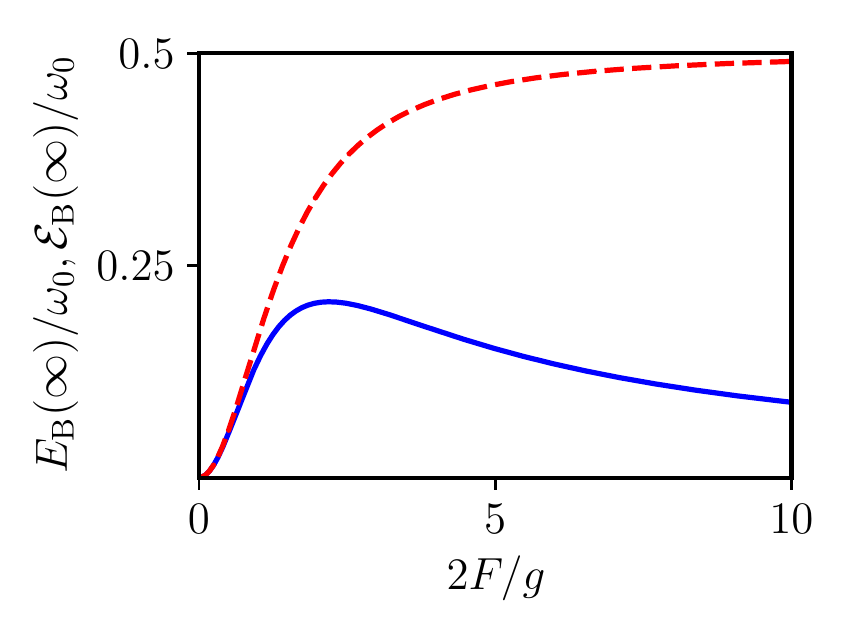}\put(0,75){(a)}\end{overpic}\vspace{0.5em}
\begin{overpic}[width=0.8\columnwidth]{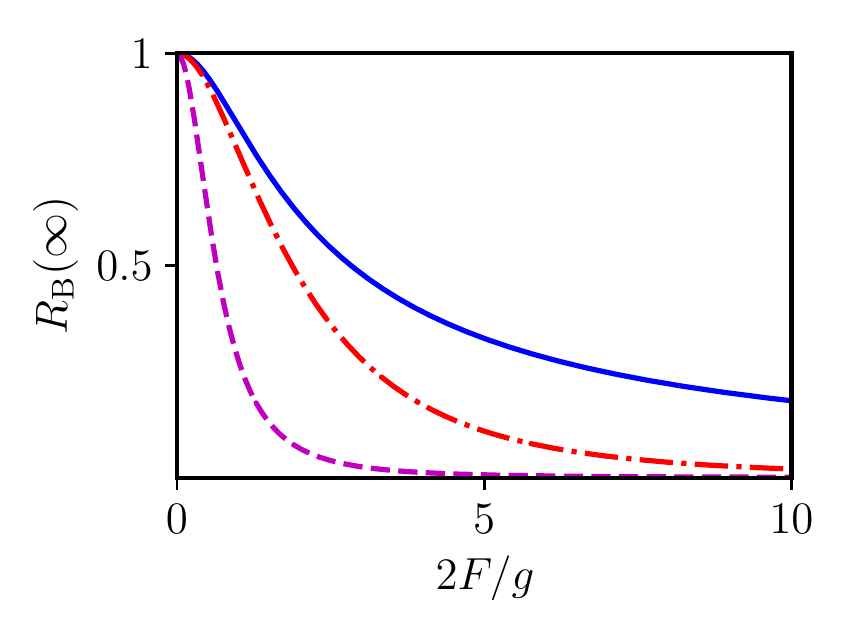}\put(0,75){(b)}\end{overpic}\vspace{0.5em}
\caption{(Color online) Panel (a) $E_{\rm B}(\infty)/\omega_{0}$ (red dashed line)---Eq.~(\ref{DERGO})---and ${\cal E}_{\rm B}(\infty)/\omega_{0}$ (blue solid line)---Eq.~(\ref{DARGO})---as functions of $2F/g$, for the two-qubit model. Results in this panel have been obtained by setting $\gamma \gg F,g$. Panel (b) The ratio $R_{\rm B}(\infty)$---Eq.~(\ref{ratio}) in the $\tau \to \infty$ limit---is plotted as a function of $2F/g$. Different curves correspond to different values of $\gamma$. Blue solid line: $\gamma \gg F,g$; red dash-dotted line: $\gamma/g = 5$; magenta dashed: $\gamma/g = 1$. Both panels refer to the purely coherent energy supply regime, i.e.~$N_{b}(T)=0$.
\label{PLOTERGOASYMPCOE}}
\end{figure}

As anticipated,   no ergotropy can be generated by only having access to a purely thermal source. Indeed, for  $F =0$, Eqs.~(\ref{EASYMP1}) and (\ref{ERGOASYMP1}) give
\begin{equation} 
{E}_{\rm B}(\infty)\Big\rvert_{F=0, T}=\omega_0 N_f(T)~, 
\end{equation}
\begin{equation}
\mathcal{E}_{\rm B}(\infty)\Big\rvert_{F=0, T}=0~,
\end{equation}
where now 
 \begin{eqnarray}  \label{NFER} 
 {N}_{f}(T)\equiv \frac{1}{\exp{[\omega_{0}/(k_{\rm B} T)]}+1}~,
 \end{eqnarray}
is the fermionic occupation number. In the opposite regime, i.e. when the charging is purely coherent and the bath is at temperature $T=0$, Eqs.~(\ref{EASYMP1}) and (\ref{ERGOASYMP1}) yield
\begin{equation}\label{DARGO}
E_{\rm B}(\infty)\Big\rvert_{F, T=0}=\omega_0\frac{(\gamma^2+8 F^2)F^2}{ 16 F^4 +\gamma^2 (2 F^2+ g^2)}~,
\end{equation}
\begin{equation} \label{DERGO}
\mathcal{E}_{\rm B}(\infty)\Big\rvert_{F, T=0}= \frac{\omega_0}{2 }\frac{g \gamma^2(\sqrt{4 F^2+g^2}-g)}{16 F^4 +\gamma^2 (2 F^2+ g^2)}~,
\end{equation}
which we plot in Fig.~\ref{PLOTERGOASYMPCOE} together with their 
ratio~(\ref{ratio}),
\begin{equation} \label{RERGO}
{R}_{\rm B}(\infty)\Big\rvert_{F, T=0}=
\frac{g\gamma^2(\sqrt{4 F^2+g^2}-g)}{2(\gamma^2+8 F^2)F^2}~.
\end{equation}
As simple analysis of Eq.~(\ref{DERGO}) reveals that 
in the large loss limit $\gamma \gg F$  and  when $F$ and $g$ are tuned so that 
 $F=  \sqrt{(\sqrt{2}+1)/2} g\simeq 1.09 g$,
 the asymptotic ergotropy reaches its maximum  value
\begin{eqnarray}  
\overline{\mathcal{E}}_{B}(\infty)\Big\rvert_{F, T=0}&=& \frac{\sqrt{2}-1}{2}\omega_0 \sim 0.207 \omega_0~, \label{MAXE} 
\end{eqnarray}
which, incidentally, corresponds also to the absolute maximum of (\ref{ERGOASYMP1})
for arbitrary $T$, as evident from panel a) of Fig.~\ref{PLOTERGOASYMPMIX}. 
On the contrary, a close inspection of Eq.~(\ref{RERGO}) reveals that the ratio achieves its absolute maximum value 1
in the small driving constant/low energy supply limit (i.e.~for $F\ll g,\gamma$) for which one gets
$E_{\rm B}(\infty)\rvert_{F, T=0} \simeq \mathcal{E}_{\rm B}(\infty)\rvert_{F, T=0} \simeq \omega_0 (F/g)^2$.
As anticipated at the beginning of this Section, this exactly reproduces the behaviour~(\ref{IMPO111}) observed for the two-harmonic oscillator model at zero temperature.

\subsubsection{Transients} \label{sec:trans}

We now analyze the perfomances of the model for finite values of $\tau$.

\begin{figure}
\begin{overpic}[width=.8\linewidth]{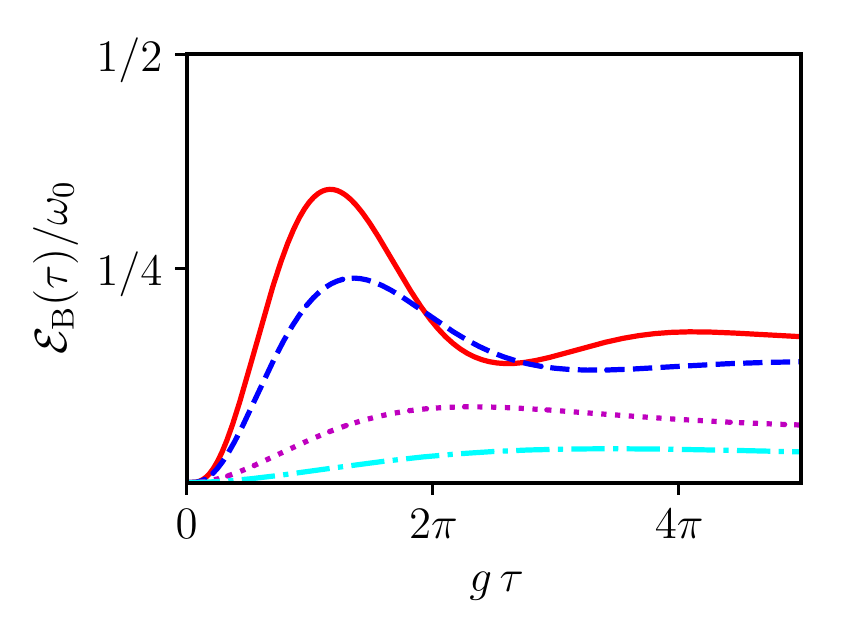}\end{overpic}\vspace{0.5em}
\caption{(Color online) ${\cal E}_{\rm B}$ (in units of $\omega_{0}$) as a function of $g\tau$, for the two-qubit model.
Different curves refer to different values of $N_{b}(T)$. Red solid line: $N_{b}(T)=0$; blue dashed line: $N_{b}(T)=0.1$; magenta dotted line: $N_{b}(T)=0.5$; cyan dash-dotted line: $N_{b}(T)=1$. Numerical results in this plot have been obtained by setting $g=F= \omega_0/5$ and $\gamma=\omega_0$. 
Notice that, in a finite range of values of $g\tau$, the result for $N_{b}(T)=0.1$ (blue dashed line) lies above the result for  $N_{b}(T)=0$ (red solid line).
\label{NEWNEW}}
\end{figure}

Let us first consider the case where
 no driving is at play ($F=0$) while the temperature of the bath is finite ($N_{b}(T)>0$) which 
 is the only case for which we can present explicit analytical expressions. 
 As for the case of the two-harmonic oscillator model, see Eq.~(\ref{DECERGOTHER}),
 it turns out that the ergotropy of the battery is always null at all times,
 i.e.~$\mathcal{E}_{\rm B}(\tau)=0$,
  testifying that  in the absence of the external driving the density matrix $\tilde{\rho}_{\rm B}(\tau)$ is passive. 
 Regarding the mean  energy of B, by direct integration of the equation of motion 
we find
\begin{eqnarray}\label{eq:analytical_battery_thermal_protocol}
 E_{\rm B}(\tau)&=&  \omega_0 {N}_{f}(T)\Big\{1+  (e^{-\frac{1}{2}{\Gamma} \tau }/\epsilon^2)  \\ \nonumber &\times & \left[{{\Gamma}^2}- \epsilon^2  -{\Gamma}{\epsilon} \sinh(\epsilon \tau / 2) - {{\Gamma}^2}\cosh(\epsilon \tau / 2) \right]\Big\}~,
\end{eqnarray}
with  $\Gamma$ and ${N}_{f}(T)$ as in Eqs.~(\ref{DEFGAMMA}) and (\ref{NFER}) respectively, and 
where
\begin{eqnarray} \label{eq:qubqub-loss-ren}
\epsilon \equiv \sqrt{{\Gamma}^2-(4 g)^2}~.
\end{eqnarray}
For comparison, we also report the value of the local mean energy of A, which in the present case is given by
\begin{eqnarray}\label{eq:analytical_charger_thermal_protocol}
E_{\rm A}(\tau)&= & \omega_0 {N}_{f}(T)\Big\{1+  (e^{-\frac{1}{2}{\Gamma} \tau }/ \epsilon^2)\\ \nonumber &\times & \left[{{\Gamma}^2}-{\epsilon^2}  + {{\Gamma}}{\epsilon} \sinh(\epsilon \tau / 2) -{\Gamma}^2\cosh(\epsilon \tau / 2) \right]\Big\}~.
\end{eqnarray}
One may notice that these expressions can be formally obtained from Eqs.~(\ref{eq:hoho-EaEb_II})
and  (\ref{eq:hoho-EaEb_I}), which apply for the two-harmonic oscillator model in the purely thermal setting (i.e.~$F=0$), by 
 replacing $N_{b}(T)\rightarrow N_{f}(T)$, $16 g^2 \rightarrow {{\Gamma}^2}-{\epsilon^2}$, and
 $\gamma \rightarrow \Gamma$.  Accordingly, in this regime the energy charging of the two-qubit model will closely resemble the one observed in Fig.~\ref{FIGnew3}, with an overdamped  and underdamped
 regime, attained respectively for ${\Gamma} \geq 4 g$ and ${\Gamma} <  4 g$, the main
 difference being that now, because of Eq.~(\ref{DEFGAMMA}), the critical threshold depends explicitly upon the bath temperature $T$.
\begin{figure*}
\centering
\begin{tabular}{cc}
\begin{overpic}[width=0.8\columnwidth]{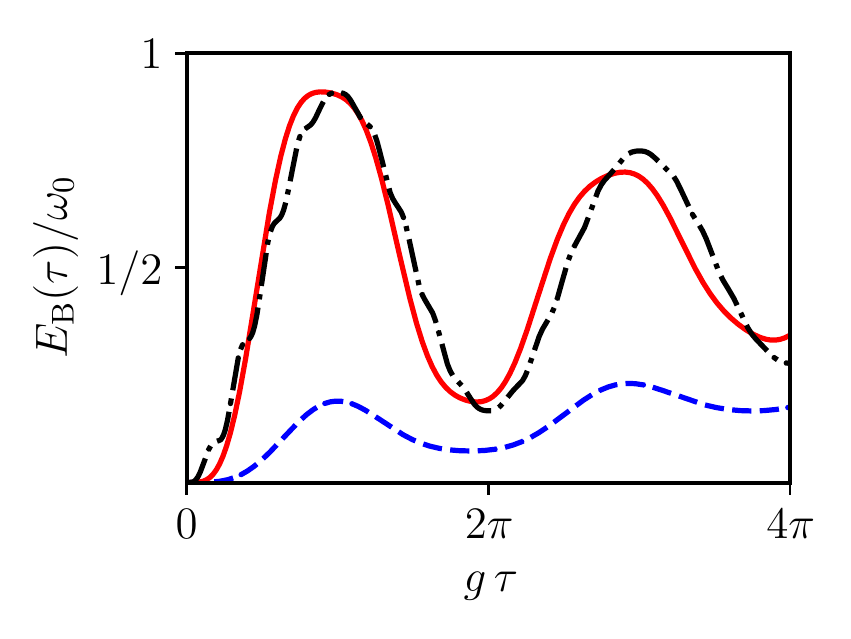}\put(10,75){\normalsize (a)}\end{overpic}\vspace{0.5em} & 
\begin{overpic}[width=0.8\columnwidth]{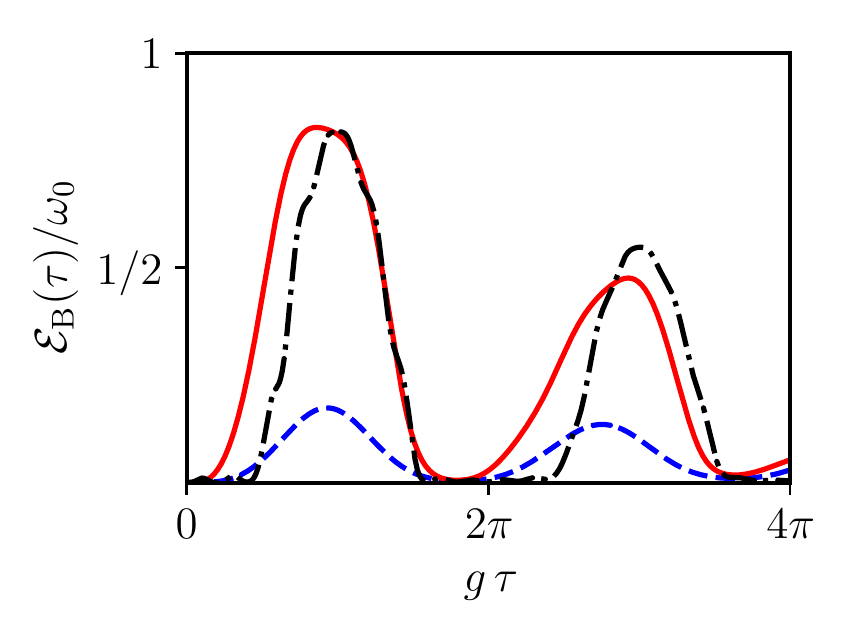}\put(10,75){\normalsize (b)}\end{overpic}\vspace{0.5em} \\
\begin{overpic}[width=0.8\columnwidth]{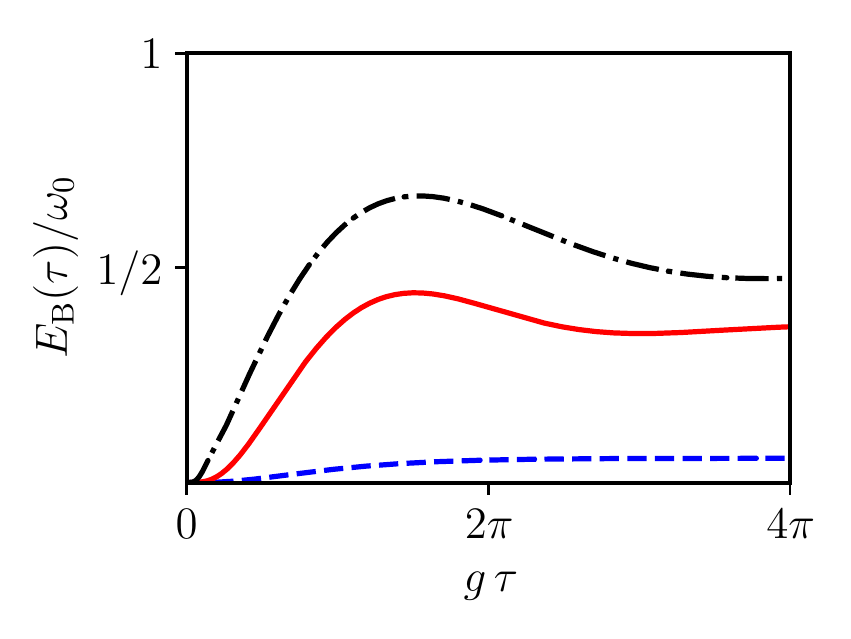}\put(10,75){\normalsize (c)}\end{overpic}\vspace{0.5em} & 
\begin{overpic}[width=0.8\columnwidth]{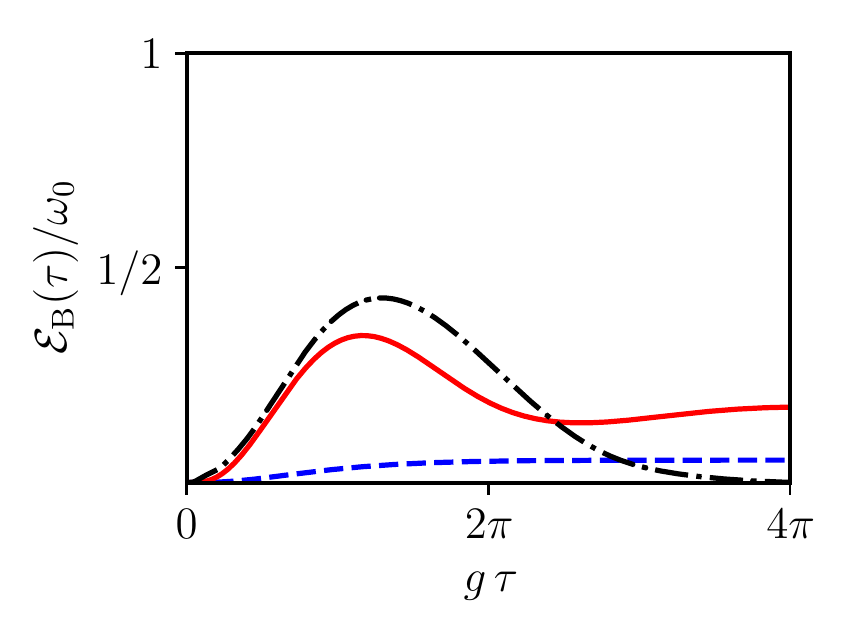}\put(10,75){\normalsize (d)}\end{overpic}\vspace{0.5em}
\end{tabular}
\caption{(Color online) Panel (a) $E_{\rm B}(\tau)$ (in units of $\omega_{0}$) as a function of $g\tau$, for the two-qubit model. Different curves refer to different values of $F$ (in units of $\omega_{0}$). Blue dashed line: $F= 0.05\omega_{0} $; red solid line: $F=0.2\omega_{0} $; black dash-dotted: $F=\omega_{0} $. 
Panel (b) Same as in panel (a) but for ${\cal E}_{\rm B}(\tau)$. Numerical results in panels (a)-(b) have been obtained by setting $g=0.2\omega_0$ and $\gamma=0.05\omega_0$. Panel (c) and (d) Same as in panels (a) and (b) but for $\gamma=\omega_0$. All results in this figure refer to the purely coherent energy supply regime, i.e.~$N_{b}(T)=0$.
\label{fig:qubqub-hoho-coh}}
\end{figure*}
To study the finite-time behaviour of $E_{\rm B}(\tau)$ and $\mathcal{E}_{\rm B}(\tau)$
 in the case where $F$ is non-zero, we resort to numerical calculations.
 In particular, in Fig.~\ref{fig:qubqub-hoho-coh} we present plots of these quantities for $T=0$ 
 (no thermal supply) 
 obtained for different values of 
 $F$, $g$, and $\gamma$. In  Fig.~\ref{NEWNEW}, instead, a study of $\mathcal{E}_{\rm B}(\tau)$ is presented for 
 fixed $F$ and various values of $N_b(T)$. Again, oscillatory behaviours can be observed which may lead to 
 an increase of $\mathcal{E}_{\rm B}(\tau)$ as a function of $T$.
 
 We conclude this Section by commenting about optimal charging times (either for $E_{\rm B}(\tau)$ or 
 $\mathcal{E}_{\rm B}(\tau)$) which, for future reference, we study in the limit of 
 strong coherent driving  ($F\gg g$) and for weak dissipation $\gamma\simeq 0$. In this limit, simple analytical solutions can be found, which for the mean energy results in 
$E_{\rm B}(\tau)=\omega_0 \sin^2(g \tau /2)$, indicating an optimal charging time 
$\pi/g$ that is independent of $F$.
\section{Hybrid model} 
\label{SEC:TRE}
The last model we consider assumes A to be a harmonic oscillator and B a qubit whose energy gap matches the frequency $\omega_0$ of A. 
Accordingly, the system  Hamiltonian is taken to be 
\begin{eqnarray}\label{eq:H-ho-qub-open} 
&{H}_{\rm A}= \omega_0 a^\dagger a~, \qquad 
{H}_{\rm B} =  {\displaystyle \frac{\omega_0}{2}} \left( \sigma_{\rm B}^z + 1\right)~, & \nonumber\\
&\Delta{H}_{\rm A}(t)= F \left( e^{-i \omega_0 t } a^\dagger + e^{i \omega_0 t } a \right)~, &\nonumber  \\ 
&{H}_{\rm AB}^{(1)}= g \left(a\, \sigma^+_{\rm B} + a^\dagger\, \sigma^-_{\rm B}\right)~,&
\end{eqnarray}
while the dissipator ${\cal D}_{\rm A}^{(T)}$ is the same we used in Sec.~\ref{subsec:hoho-open}, i.e. it is provided 
by Eq.~(\ref{eq:me-qub-qub111}). Switching to the interaction picture representation, the resulting ME~(\ref{eq:me-qub-qubSCHEMAfree}) is hence given  by 
\begin{eqnarray}
&&\dot{\tilde{\rho}}_{\rm A B}(t)=-i \left[g \left(a\, \sigma^+_{\rm B} + a^\dagger\, \sigma^-_{\rm B} \right)+F (a^\dagger +   a),\; \tilde{\rho}_{\rm A B}(t) \right]\nonumber \\
&&\;+ \gamma(N_{b}(T)+1)  \mathcal{D}_{\rm A}^{[a]}\left[ \tilde{\rho}_{\rm A B}(t)  \right]
+ \gamma N_{b}(T) \mathcal{D}_{\rm A}^{[a^\dag]} \left[\tilde{\rho}_{\rm A B}(t)  \right]~. \nonumber \\ \label{eq:at-ho-lindblad-me}
\end{eqnarray}
Being the system hybrid and infinite-dimensional, the integration methods adopted  in the previous two cases cannot be applied
as they will produce an infinite set of coupled differential equations.
Instead, we resort to the 
characteristic function approach~\cite{WallsMilburn2007, Lougovski07,GAUSSIANCV}, which allows one to cast Eq.~(\ref{eq:at-ho-lindblad-me}) into a finite
set of linear partial differential equations that can be solved numerically. 
 By choosing this approach, we pass from infinite square matrices (density matrix formalism) to four complex functions for describing the system's state.
For this purpose, we decompose $\tilde{\rho}_{\rm A B}(t)$ into the basis of the energy eigenstates $\{ |0\rangle_{\rm B}, |1\rangle_{\rm B}\}$, of $H_{\rm B}$, i.e.
\begin{eqnarray}\label{eq:at-ho-rho-dec}
\tilde{\rho}_{\rm AB}(t)&=&\sum_{i j} \tilde{\rho}_{\rm A}^{(ij)}(t) \otimes \ket{i}_{\rm B}\!\bra{j}~. 
\end{eqnarray}
Here, 
$\tilde{\rho}_{\rm A}^{(ij)}(t) \equiv {_{\rm B}\!}\bra{i} \tilde{\rho}_{\rm AB}(t) \ket{j}_{\rm B}$
are operators of A which we express as a convolution integral,
\begin{eqnarray}
\tilde{\rho}_{\rm A}^{(ij)}(t) = \int  \frac{d^{2} \beta}{\pi} \chi_{ij}(\beta,t) D(-\beta)~, \label{INVE}
\end{eqnarray} 
over a complex variable $\beta$ of the displacement operator $D(\beta)\equiv \exp\left(\beta a^\dagger - \beta^* a \right)$ and
\begin{equation}\label{CHI} 
\chi_{ij}(\beta,t)\equiv {\rm tr}_{\rm A} \left[ D(\beta) \tilde{\rho}_{\rm A}^{(ij)}(t) \right]~,
\end{equation}
where the latter quantity is the  associated (symmetrically ordered) characteristic $\chi$-function~\cite{WallsMilburn2007}. They inherit from $\tilde{\rho}_{\rm AB}(t)$ the 
following constraints
\begin{eqnarray}
&&\chi_{00}(0,t)+\chi_{11}(0,t) =1~,
\label{eq:chi-trace-condition} \\ 
&&\chi_{ij}(\beta,t)=\chi_{ji}^*(-\beta,t)~,
\label{eq:chi-selfadjointness}
\end{eqnarray}
the first deriving from the normalization of  $\tilde{\rho}_{\rm AB}(t)$, the second from its self-adjointness. Furthermore, considering that B is a qubit so that  Eqs.~(\ref{stored energy1}) and (\ref{eq:ergo1}) can be still exploited, Eq.~(\ref{eq:ergo1}) allows us to express the quantity of interest in the following compact form
\begin{eqnarray}
{E_{\rm B}(\tau)}&=&
 {\omega_0} \chi_{11}(0,\tau)~,  \\  \nonumber 
\mathcal{E}_{\rm B}(\tau)&=& {\displaystyle \frac{\omega_0}{2}}
\left[\sqrt{(\chi_{11}-\chi_{00})^2+4|\chi_{10}|^2}
+ \chi_{11}-\chi_{00} \right]\Big\rvert_{\beta=0}~.
\end{eqnarray}
Replacing Eq.~(\ref{CHI}) into  (\ref{eq:at-ho-rho-dec})  and exploiting the algebra of the harmonic oscillator, we can now recast the ME~(\ref{eq:at-ho-lindblad-me}) 
  into a set of partial differential equation for $\chi_{ij}(\beta,t)$, i.e. 
\begin{eqnarray}
\label{eq:chi}
\dot{\chi}_{ij} &=& - i g \mathcal{I}_{ij}\left[\vec{\chi}\right]+2 i F x \chi_{ij}
\\ \nonumber 
&&-  \gamma  \left[ \left(N_{b}(T)+\frac{1}{2}\right)\left(x^2+y^2 \right)+\frac{1}{2}\left( x \partial_x + y \partial_y\right)\right] \chi_{ij} ~,
\end{eqnarray}
where $x$ and $y$ are the real and imaginary components of $\beta= x+iy$ and $\mathcal{I}_{ij}\left[\vec{\chi}\right]$ are differential terms 
describing the energy exchange between the harmonic oscillator and the qubit, 
\begin{widetext}
\begin{equation}
\begin{cases}
 \mathcal{I}_{11}\left[ \vec{\chi} \right]= -\frac{1}{2}\left[ 
\left( \partial_x - i \partial_y \right)\chi_{10} + \left( \partial_x + i \partial_y \right)\chi_{01} + \left( x - i y \right)\chi_{10} + \left( x + i y\right)\chi_{01} \right] ~\nonumber\\
 \mathcal{I}_{10}\left[ \vec{\chi}\right]= -\frac{1}{2}\left[ 
\left( \partial_x + i \partial_y \right)\left(\chi_{00}-\chi_{11} \right)
+
\left( x + i y\right)\left(\chi_{11}+\chi_{00} \right) 
 \right]~\nonumber \\
  \mathcal{I}_{01}\left[ \vec{\chi}\right]= \frac{1}{2}\left[ 
\left( \partial_x - i \partial_y \right)\left(\chi_{11}-\chi_{00} \right)
-
\left( x - i y\right)\left(\chi_{11}+\chi_{00} \right) 
 \right]~\nonumber \\ 
  \mathcal{I}_{00}\left[ \vec{\chi} \right]= \frac{1}{2}\left[ 
\left( \partial_x - i \partial_y \right)\chi_{10} + \left( \partial_x + i \partial_y \right)\chi_{01} -\left( \left( x - i y \right)\chi_{10} + \left( x + i y\right)\chi_{01} \right)\right] ~.
\end{cases}
\end{equation}
\end{widetext}
(It is worth noticing that the set~(\ref{eq:chi}) embodies both the constraints of Eqs.~(\ref{eq:chi-trace-condition}) and (\ref{eq:chi-selfadjointness}).)

\begin{figure*}
\centering
\begin{tabular}{cc}
\begin{overpic}[width=0.8\columnwidth]{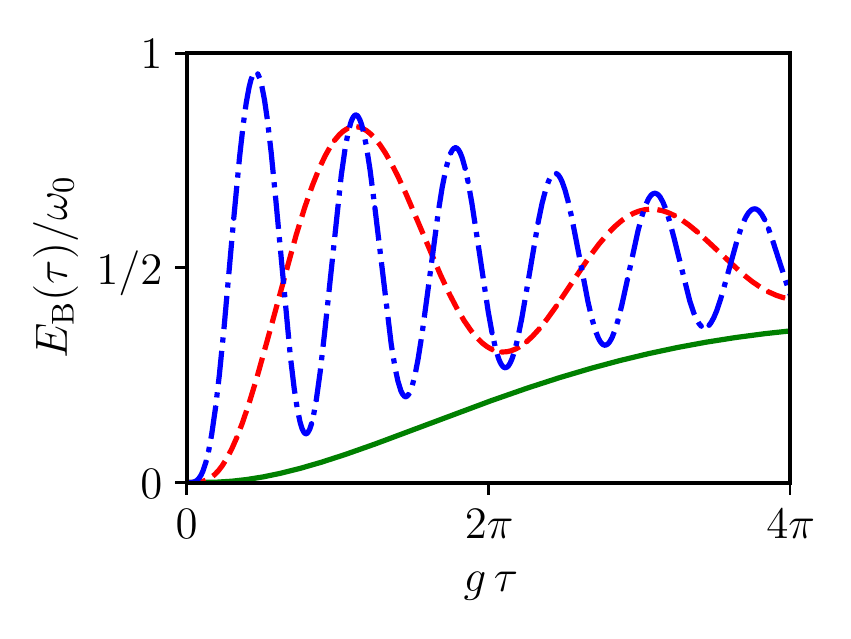}\put(10,75){\normalsize (a)}\end{overpic}\vspace{0.5em} & 
\begin{overpic}[width=0.8\columnwidth]{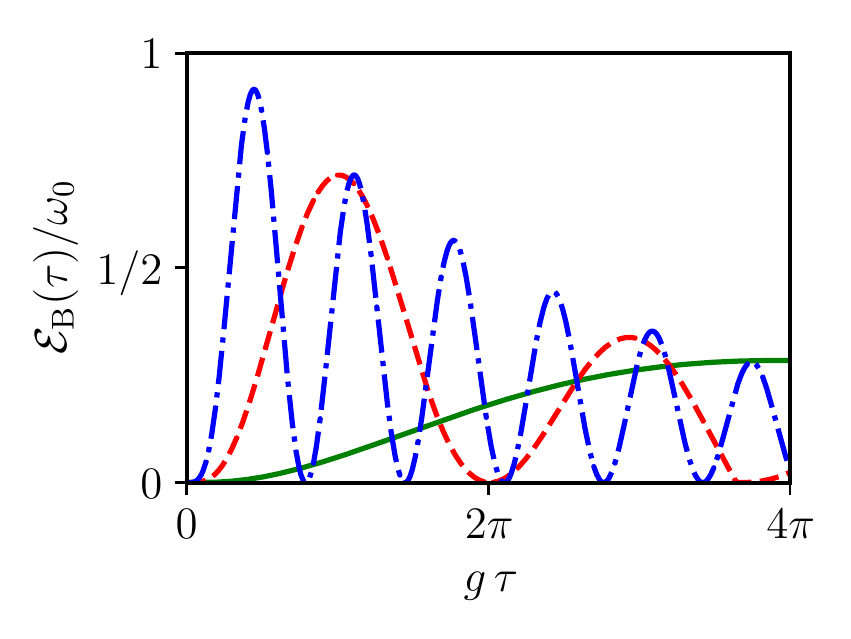}\put(10,75){\normalsize (b)}\end{overpic}\vspace{0.5em} \\
\begin{overpic}[width=0.8\columnwidth]{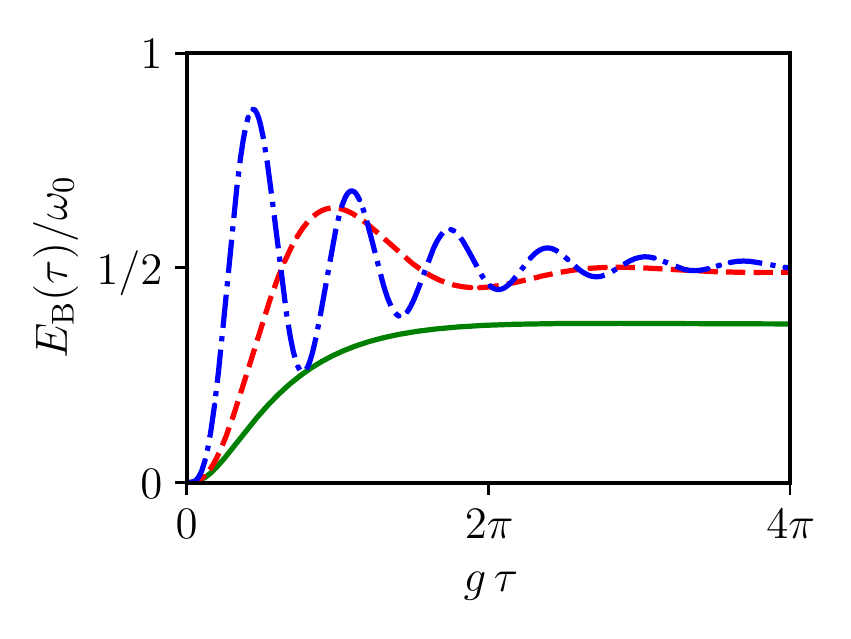}\put(10,75){\normalsize (c)}\end{overpic}\vspace{0.5em} & 
\begin{overpic}[width=0.8\columnwidth]{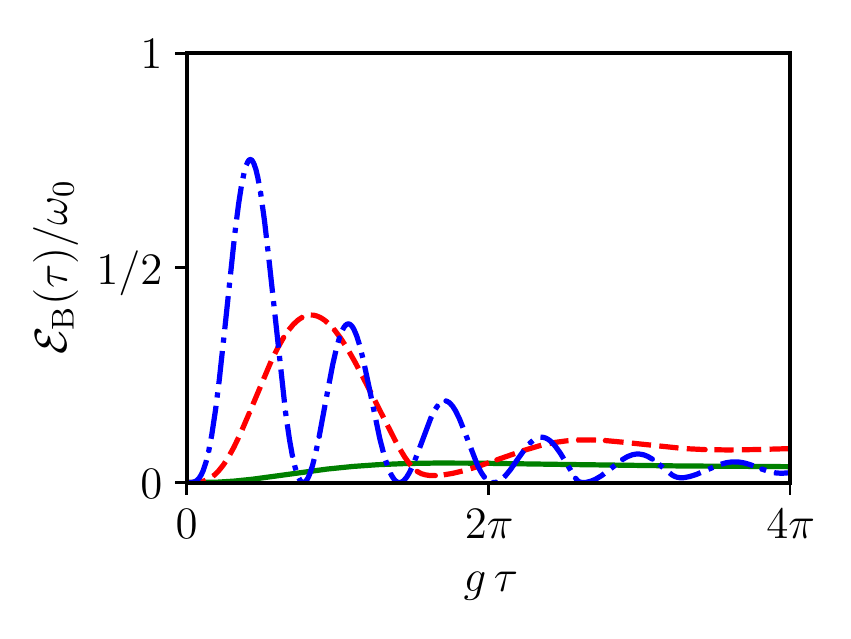}\put(10,75){\normalsize (d)}\end{overpic}\vspace{0.5em}
\end{tabular}
\caption{(Color online)
Panel (a) $E_{\rm B}(\tau)$ (in units of $\omega_{0}$) as a function of $g\tau$, for the hybrid model. Different curves refer 
to different values of $F$ (in units of $\omega_{0}$). Green solid line: $F= 0.1\omega_{0} $; red dashed line: $F=0.5\omega_{0} $; blue dash-dotted line: $F =1.5\omega_{0}$. 
Panel (b) Same as in panel (a) but for ${\cal E}_{\rm B}(\tau)$.
Numerical results in panels (a)-(b) have been obtained by setting $g=0.1\omega_0$, $\gamma=\omega_0$, and $N_{b}(T)=0$.
Panels (c) and (d) Same as in panels (a) and (b) but for $N_{b}(T)=1$.\label{fig:hoqub-coh-Es}}
\end{figure*}

\begin{figure}
\begin{overpic}[width=0.8\linewidth]{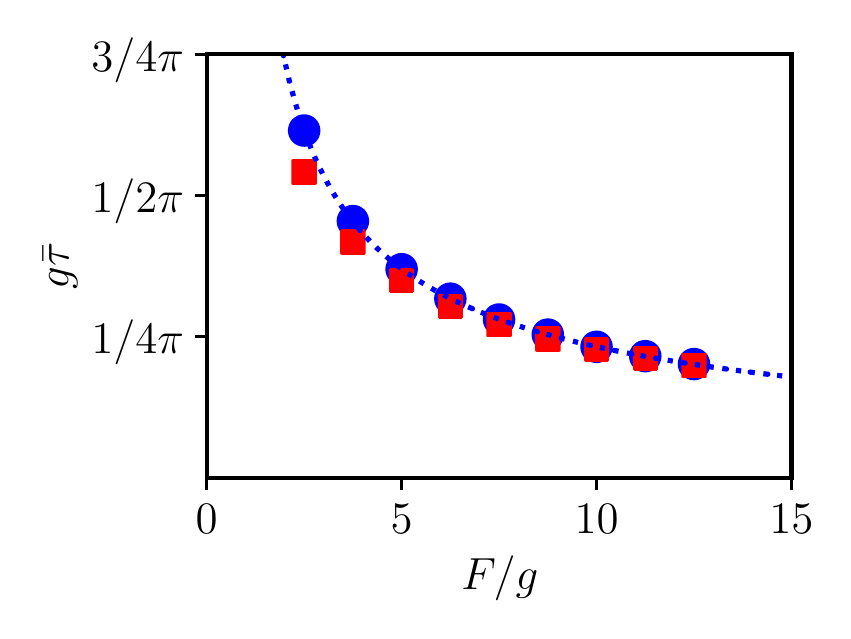}\end{overpic} 
\caption{(Color online) 
The time scale $\bar{\tau}$ (in units of $1/g$) at which $E_{\rm B}(\tau)$ reaches its maximum value (blue circles) is plotted as a function of $F/g$. Red squares denote the same quantity but for the case of $\mathcal{E}_{\rm B}(\tau)$. Both results refer to the hybrid model. Numerical results in this figure have been obtained by setting
$g=0.2 \omega_0$, $\gamma=\omega_0$, and refer to the purely coherent energy supply regime, i.e.~$N_{b}(T)=0$.
\label{fig:timefit}}
\end{figure}

Equations (\ref{eq:chi}) have been solved numerically under the usual initial conditions~(\ref{eq:in-cond-qubqub}), which, casted into the $\chi$-function language, read
\begin{eqnarray} \label{eq:chi-in-cond}
\chi_{00}(\beta,0)&=&e^{-\frac{|\beta|}{2}^2}~,\nonumber \\
\chi_{11}(\beta,0)&=&\chi_{10}(\beta,0)=\chi_{01}(\beta,0)=0~. 
\end{eqnarray}
For the case where $F=0$ (no coherent driving) our findings are in agreement with the two previous cases. 
Specifically, no ergotropy on B is generated, while, regarding $E_{\rm B}(\tau)$,  
for small values of $\gamma/g$ an oscillating behaviour is observed which is then lost for large $\gamma/g$,
 the thermalization value  being 
  $E_{\rm B}(\infty)=\omega_0  N_{f}{(T)}$ (data not shown). 
As we turn on $F$, non-zero values of $\mathcal{E}_{\rm B}(\tau)$ are observed with an oscillatory behaviour that reminds us of the results of the previous section,  see Fig.~\ref{fig:hoqub-coh-Es}. 
By numerical analysis we also study the optimal charging times 
(see Fig.~\ref{fig:timefit}) noticing that for the hybrid model they appear to have a 
 $1/\sqrt{F}$  scaling, at least for  large values of the field amplitude $F$. 
This is deeply different with respect to the two-qubit case for which a finite charging time emerges in the same regime, and also with respect to the case of two harmonic oscillators, where the driving amplitude $F$ does not enter in the time scales of the charging process. This peculiarity is a consequence of the structure of the Hilbert space of the hybrid system studied in this Section. Indeed, the quantum harmonic oscillator A can host an arbitrarily large number of excitations coming from the interaction with the  coherent source, while the qubit (i.e.~the battery B) has an upper bounded spectrum: hence, the more energy is in the mediator the lesser the charging time of the qubit is.

\section{Summary and conclusions}
\label{Conclusions}

In this Article we have studied charger-mediated energy transfer for quantum batteries via an open system approach. Specifically, we have studied three models. One in which both the charger A and the quantum battery B are 
described by harmonic oscillators, described in Sect.~\ref{subsec:hoho-open}, one in which both A and B are qubits, described in Sect.~\ref{sect:open-TLS}, and, finally, one in which A is a harmonic oscillator and B is a qubit, see Sect.~\ref{SEC:TRE}. In all cases, the charger A interacts with an external energy supply E, and acts as mediator between E and B. At the beginning of the charging protocol, both A and B are in the ground state with zero energy, and energy is dynamically injected into the system thanks to the presence of E, either via the presence of a thermal bath at temperature $T$ or via a coherent driving field of amplitude $F$. Particular attention has been devoted to the maximum extractable work from B, i.e.~the so-called ergotropy.

Our main findings can be summarized as following.
(i) The case of two harmonic oscillators is profoundly different from the other two cases. Because of the linearity of the system, in the case of two harmonic oscillators there is no interplay between the coherent and incoherent energy supplies. In particular, in the coherent protocol ($F > 0$, $T = 0$), ergotropy and energy coincide. This happens because A and B remain uncorrelated during the system's evolution.
(ii) In the case of the thermal protocol ($F = 0$, $T > 0$), the ergotropy is always zero. This holds true for all models.
(iii) In the case of two qubits in the mixed regime ($F > 0$, $T > 0$) (while typically non-zero temperature tends to reduce the ergotropy) there are special settings for which finite temperature is beneficial for the ergotropy. This is a consequence of the nonlinear character of this model, which leads to a non-trivial interplay among coherent and incoherent channels. 
(iv) In the hybrid model, the time at which energy and ergotropy are maximal decreases monotonically with the driving field $F$. This peculiarity stems from the structure of the Hilbert space of the hybrid model and can be compared with the energy dynamics derived in Ref.~\onlinecite{ALTRO}, in a closed (i.e.~Hamiltonian) setting.

\acknowledgments
We gratefully thank M. Campisi, F.M.D. Pellegrino, D. Ferraro, P.A. Erdman, V. Cavina, and M. Keck for useful discussions.

\appendix

\section{Details on the ergotropy functional} \label{ERGOGO}

Let $\rho$ be the density matrix of a system characterized by a Hamiltonian $H$, 
which we present in terms of their spectral decompositions:
 \begin{eqnarray} \rho&=&\sum_{n} r_n |r_n\rangle\langle r_n|~,  \\
H&=& \sum_{n} e_n |e_n\rangle\langle e_n|~. 
\end{eqnarray}
Here, $\{ |r_n\rangle\}_n$ and $\{|e_n\rangle\}_n$ represent the eigenvectors
 of $\rho$ and $H$, respectively, and $r_0\geq r_1\geq \cdots$
 and $\epsilon_0 \leq \epsilon_1 \leq \cdots$ are the associated eigenvalues, which we have been properly ordered.
 The passive counterpart of $\rho$ is defined as the following density matrix~\cite{ERGO,PASSIVE} 
 \begin{equation} 
 \rho^{(p)} \equiv \sum_{n} r_n |e_n\rangle\langle e_n|~.
 \end{equation}
 By construction, its mean energy is given by
\begin{equation} 
E^{(p)} \equiv {\rm tr}[{H} \rho^{(p)}]= \sum_n r_n \epsilon_n~,
\end{equation}
 and, as mentioned in Sect.~\ref{SEC:FIGs}, it corresponds to the last term in the r.h.s. of Eq.~(\ref{eq:ergo}), i.e.
\begin{equation}    
E^{(p)}=\min_{\substack{U}}{\rm tr}\left[ {H} U \rho  U^\dagger \right]~.
\end{equation}
Accordingly, the ergotropy $\mathcal{E}$ of the state $\rho$ can be conveniently expressed as
\begin{equation}
\mathcal{E} = E  -E^{(p)} = {\rm tr}[{H} (\rho- \rho^{(p)})]~,\label{eq:ergoEQUI}
\end{equation}
which makes it evident that zero values of $\mathcal{E}$ can be obtained only for those density matrices
which are passive, i.e.~for $\rho= \rho^{(p)}$. 
From the above construction  it is also clear that states differing by a unitary transformation $V$ 
(e.g.~$\rho$ and $\rho'=V \rho V^\dag$) will have the same 
passive state. Accordingly, we can write the ergotropy of $\rho'$ as
\begin{equation}
\mathcal{E}' =  {\rm tr}[{H} (\rho'- \rho^{(p)})]  = {E}'-{E}+  \mathcal{E}~,\label{eq:ergoEQUI1}
\end{equation}
with ${E}=\mbox{tr}[{H} \rho]$ and ${E}'=\mbox{tr}[{H} \rho']$ the mean energies of $\rho$ and  $\rho'$, respectively.

Exploiting the above identities we can easily produce closed-form expressions for the ergotropy of special cases.
Consider the case of a qubit with an Hamiltonian of the form 
${H} = \omega_{0}(\sigma^z+1)/2$ and density matrix
\begin{equation}
{\rho}=\frac{1}{2}(\openone+ \vec{r} \cdot \vec{\sigma})~,
\label{eq:ergo-qubqubapp}
\end{equation}
where $ \openone$ is the $2 \times 2$ identity and $\vec{\sigma} \equiv (\sigma^{x},\sigma^{y},\sigma^{z})$ and $\vec{r}$ are the Pauli and Bloch vectors, respectively.
Then simple algebric manipulations yield
\begin{equation}
\mathcal{E}=\frac{\omega_0}{2}\left({r}+r_z\right)~,\label{eq:ergo1IMPO} 
\end{equation}
with $r=|\vec{r}|$. Eq.~(\ref{eq:ergo1IMPO}) can be written in terms of expectation values of operators as
\begin{equation}
\mathcal{E} = \frac{\omega_0}{2}\left(\sqrt{\left<\sigma_z\right>^2 + 4 \left<\sigma_+\right> \left<\sigma_-\right>}+ \left<\sigma_z\right>\right)~.\label{eq:ergo-qubqub-lang} 
\end{equation}
With reference to Sect.~\ref{sect:open-TLS}, this form shows that when the energy is supplied only thermally, i.e.~$F=0$ and $N_{b}(T)>0$, the ergotropy is null because $\left<\sigma_z\right><0$ and $\left<\sigma_+\right>\left<\sigma_-\right>=0$ for all the time evolution.

We now discuss the case of two harmonic oscillators, see Sect.~\ref{subsec:hoho-open}.
In this case, a closed-form expression for the ergotropy can be derived by noticing that both the state of the system AB and the reduced states of A and B are Gaussian for all values of $\tau$. Since we want to extract energy from B, from now on we just concentrate on the reduced density matrix of the quantum battery B. A Gaussian state $\rho_{\rm G}$ can be obtained from a thermal state $\rho_{\bar{\beta}} \propto \exp(-{\bar{\beta}} H_{\rm B}) $ of inverse temperature ${\bar{\beta}}$ by using the following identity~\cite{GAUSSIANCV}:
\begin{equation}
\rho_{\rm G}=   D^\dagger(\alpha) S^\dagger (\xi) \rho_{\bar{\beta}} S (\xi) D(\alpha)~,
\label{eq:gaussian-state}
\end{equation}
where $D (\alpha)=\exp(\alpha b^\dagger - \alpha^* b)$ and 
$S (\xi)=\exp[(\xi^* b^2 - \xi b^{\dagger \, 2})/2]$ are displacement and squeezing operators, respectively. Here, $\alpha$ and $\xi$ are the displacement and squeezing complex parameters~\cite{GAUSSIANCV} that identify the Gaussian state, together with the real parameter $\bar{\beta}$.

We now observe that the ME (\ref{eq:ho-ho-lindblad-me}) involves at most quadratic combinations of the operators
$a,b,a^\dagger, b^\dagger$. This implies that the resulting time evolution only maps Gaussian states into other Gaussian states, ensuring that $\rho_{\rm B}(\tau)$ can be written in the form (\ref{eq:gaussian-state}).
Furthermore, noting that any thermal state is passive, the ergotropy of a Gaussian state can be written as the difference between the energy of $\rho_{\rm G}$ and the energy of the thermal state $\rho_{\bar{\beta}}$, which is connected to it via displacement and squeezing~\cite{hofer-gaussian-2018}:
\begin{equation}
\mathcal{E}[\rho_{\rm G}]=\omega_0 {\rm tr }[b^\dagger b \rho_{\rm G}]-\omega_0 {\rm tr }[b^\dagger b \rho_{\bar{\beta}}]~.
\label{eq:ergo-gaussian}
\end{equation} 
Hence, in order to calculate the ergotropy of $\rho_{\rm G}$ we need to determine the mean occupation number $n_{\bar{\beta}} = {\rm tr }[b^\dagger b \rho_{\bar{\beta}}]$ of $\rho_{\bar{\beta}}$. We introduce the canonical variables of the joint system, 
$x_{\rm A} \equiv (a + a^\dagger)/\sqrt{2}$,
$p_{\rm A} \equiv (a - a^\dagger)/(\sqrt{2} i)$,
$x_{\rm B} \equiv (b + b^\dagger)/\sqrt{2}$,
$p_{\rm B} \equiv (b - b^\dagger)/(\sqrt{2} i)$,
the vector 
$
\vec{r}=(x_{\rm A}, p_{\rm A}, x_{\rm B}, p_{\rm B})^T
$,
and the covariance matrix 
$
\sigma_{ij}\equiv {\rm tr } \left[\rho_{\rm G} \left(r_l - \left<r_l\right> \right)\left(r_m - \left<r_m\right> \right)\right]
$, whose dynamics is determined by Eqs.~(\ref{eq:hoho-lang})-(\ref{eq:hoho-lang-in-cond}), with $l, m \in \{1,2,3,4\}$. 
We also introduce $\vec{r}^{({\rm B})}=(x_{\rm B}, p_{\rm B})^T$ and the covariance matrix $\sigma_{\rm B}$ of B, which is nothing but the bottom-right $2\times 2$ block of the full covariance matrix $\sigma$. Now, $\sigma_{\rm B}$ admits the following symplectic decomposition,
\begin{equation}\label{eq:symplectic-decomposition}
\sigma_{\rm B} = (2 N_{b}(\bar{T}) +1) \mathbb{S}(\xi) \mathbb{S}^T(\xi)~,
\end{equation}
where $\mathbb{S}$ is the sympectic matrix representation of the squeezing operator in Eq.~(\ref{eq:gaussian-state}), 
$\mathbb{S}^T$ is its transpose~\cite{GAUSSIANCV}, $\bar{T}=1/(k_{\rm B}\bar{\beta})$, and the function $N_{b}(T)$ has been introduced in Eq.~(\ref{BoseEinstein}).

The matrix elements of $\mathbb{S}$ can be obtained from the identity 
$
S(\xi) r^{({\rm B})}_l S^\dagger(\xi)=\sum_m \mathbb{S}_{lm}(\xi) r^{({\rm B})}_m
$
and satisfy the symplectic group condition 
$
\mathbb{S} \Omega \mathbb{S}^{T}=\Omega~,
$
with
\begin{equation}
\Omega_{i,j}=-i [r^{({\rm B})}_i, r^{({\rm B})}_j] =
\begin{pmatrix}
0 & 1\\
-1 & 0
\end{pmatrix}~.
\end{equation}
By imposing the symplectic condition we find 
\begin{equation}
{\rm det}(\sigma_{\rm B})=(2 N_{b}(\bar{T}) +1)^2
\end{equation}
and the desired expression for the ergotropy of B:
\begin{equation}
\mathcal{E}_{\rm B}=\omega_0 \left({\rm tr }[b^\dagger b \rho_{\rm G}]-\frac{\sqrt{{\rm det}\left(\sigma_{\rm B}\right)}-1}{2}
\right)~,
\label{eq:ergo-hoho-final}
\end{equation}
with $\rho_{\rm G}=\rho_{\rm B}(\tau)$ being the state of B at a generic time $\tau$.

Finally, employing the definition of $\vec{r}^{({\rm B})}$ in terms of the creation and annihilation operators $b, b^\dagger$, one can easily write the determinant of ${\rm det}(\sigma_{\rm B})$ as in Eq.~(\ref{DEFdet}) of the main text.

\section{Energy and ergotropy decoupling for the two-oscillator model} 
\label{DECOUPLING}

In this Appendix we present an explicit proof of the decoupling described in Eq.~(\ref{DEC1}), which, for ease of notation, 
we rewrite here as
 \begin{eqnarray} \label{DEC1rew} 
 \langle x\rangle = 
  \langle x\rangle_{\rm th} + \langle x\rangle_{\rm co}~,
 \end{eqnarray}
 where we introduced  the simplified symbols $\langle x\rangle_{\rm th}\equiv \langle x\rangle\rvert_{F=0,T}$ and $\langle x\rangle_{\rm co}\equiv \langle x\rangle\rvert_{F,T=0}$. 

According to our definitions, the quantities $\langle x\rangle_{\rm th}$ for all operators $x$ appearing in 
Eqs.~(\ref{eq:hoho-lang})--(\ref{hoho-lang-3}) can be obtained by solving these equations with $F=0$, i.e.~
\begin{equation} \label{eq:hoho-lang1the}
\begin{cases}
\dot{\left\langle a\right\rangle}_{\rm th}=-i g\langle b\rangle_{\rm th} - {\displaystyle \frac{\gamma}{2}} \langle a\rangle_{\rm th}~, \\
\dot{\langle b \rangle}_{\rm th}=- i g \langle a\rangle_{\rm th}~,\\
 \dot{\langle a b^\dagger \rangle}_{\rm th}=i \left[g( \langle a^\dagger a \rangle_{\rm th} -  \langle b^\dagger b\rangle_{\rm th})\right]- {\displaystyle \frac{\gamma}{2}} \langle a b^\dagger\rangle_{\rm th}~,\\
\dot{\langle b^\dagger b \rangle}_{\rm th}=2g~{\rm Im}{\langle a b^\dagger\rangle_{\rm th}}~,\\ 
\dot{\langle a^\dagger a  \rangle}_{\rm th}= -2 g~{\rm Im} \langle a b^\dagger\rangle_{\rm th}- \gamma \langle a^\dagger a\rangle_{\rm th} + \gamma N_{b}(T)~, \\
\dot{\langle a^2\rangle}_{\rm th}=-2 i g \langle ab\rangle_{\rm th} -\gamma \langle a^2\rangle_{\rm th}~,\\
\dot{\langle ab\rangle}_{\rm th}=-ig (\langle a^2\rangle_{\rm th}+\langle b^2\rangle_{\rm th}) - {\displaystyle \frac{\gamma}{2}}\langle ab\rangle_{\rm th}~,\\
\dot{\langle b^2\rangle}_{\rm th}=-2ig \langle ab\rangle_{\rm th} \,,
\end{cases}
\end{equation}
with the initial conditions
\begin{eqnarray} 
\langle a \rangle_{\rm th}\rvert_{t=0} &=& \langle b \rangle_{\rm th}\rvert_{t=0} =0~,\nonumber\\
 \langle a^\dagger a  \rangle_{\rm th}\rvert_{t=0} &=& \langle b^\dagger b  \rangle_{\rm th}\rvert_{t=0} =
   \langle a b^\dagger \rangle_{\rm th}\rvert_{t=0}=0 \nonumber~, \\
  \langle a^2   \rangle_{\rm th}\rvert_{t=0} &=& \langle b^2   \rangle_{\rm th}\rvert_{t=0}= \langle a b   \rangle_{\rm th}\rvert_{t=0}=0~. \label{eq:hoho-lang-in-condthe}
 \end{eqnarray}
 Eqs.~(\ref{eq:hoho-lang1the})-(\ref{eq:hoho-lang-in-condthe}) imply
 \begin{eqnarray} 
 \langle a \rangle_{\rm th}&=& \langle b \rangle_{\rm th} =0\label{SOL1}~,\\
 \langle a^2 \rangle_{\rm th}&=& \langle b^2 \rangle_{\rm th} = \langle ab \rangle_{\rm th}=0~,  \label{SOL2}
 \end{eqnarray}
 at all times $t$. 
 
Similarly, the functions $\langle x\rangle_{\rm co}$ solve Eqs.~(\ref{eq:hoho-lang})-(\ref{hoho-lang-3})  with $N_{b}(T)=0$, i.e.
\begin{equation} \label{eq:hoho-lang1co}
\begin{cases}
\dot{\left\langle a\right\rangle}_{\rm co}=-i (g\langle b\rangle_{\rm co} +F)- {\displaystyle \frac{\gamma}{2}} \langle a\rangle_{\rm co}~, \\
\dot{\langle b \rangle}_{\rm co}=- i g \langle a\rangle_{\rm co}~,\\
 \dot{\langle a b^\dagger \rangle}_{\rm co}=i \left[g( \langle a^\dagger a \rangle_{\rm co} -  \langle b^\dagger b\rangle_{\rm co})- F\langle b\rangle_{\rm co}^*\right]- {\displaystyle \frac{\gamma}{2}} \langle a b^\dagger\rangle_{\rm co}~,\\
\dot{\langle b^\dagger b \rangle}_{\rm co}=2g~{\rm Im}{\langle a b^\dagger\rangle_{\rm co}}~,\\ 
\dot{\langle a^\dagger a  \rangle}_{\rm co}= -2 ~{\rm Im}(g \langle a b^\dagger\rangle_{\rm co}+F\langle a\rangle_{\rm co})- \gamma \langle a^\dagger a\rangle_{\rm co}~, \\
\dot{\langle a^2\rangle}_{\rm co}=-2 i (g \langle ab\rangle_{\rm co}+F \langle a\rangle_{\rm co}) -\gamma \langle a^2\rangle_{\rm co}~,\\
\dot{\langle ab\rangle}_{\rm co}=-i[g (\langle a^2\rangle_{\rm co}+\langle b^2\rangle_{\rm co})+F \langle b\rangle_{\rm co}] - {\displaystyle \frac{\gamma}{2}}\langle ab\rangle_{\rm co}~,\\
\dot{\langle b^2\rangle}_{\rm co}=-2ig \langle ab\rangle_{\rm co}~,
\end{cases}
\end{equation}
with initial conditions
\begin{eqnarray} 
\langle a \rangle_{\rm co}\rvert_{t=0} &=& \langle b \rangle_{\rm co}\rvert_{t=0} =0~, \nonumber\\
 \langle a^\dagger a  \rangle_{\rm co}\rvert_{t=0} &=& \langle b^\dagger b  \rangle_{\rm co}\rvert_{t=0} =
   \langle a b^\dagger \rangle_{\rm co}\rvert_{t=0}=0~,\nonumber\\
  \langle a^2   \rangle_{\rm co}\rvert_{t=0} &=& \langle b^2   \rangle_{\rm co}\rvert_{t=0}= \langle a b   \rangle_{\rm co}\rvert_{t=0}=0~. \label{eq:hoho-lang-in-condco}
 \end{eqnarray}
Eq.~(\ref{DEC1rew})---or Eq.~(\ref{DEC1}) in the main text---follows from the simple observation that the functions
$\langle x\rangle_{\rm th}+ \langle x\rangle_{\rm co}$ solve Eqs.~(\ref{eq:hoho-lang})-(\ref{hoho-lang-3}) and by using Eq.~(\ref{SOL1}).

We now demonstrate the decoupling identities~(\ref{DECERGOCOE}) and (\ref{DECERGOTHER}) for the ergotropy. 
The latter is simply a consequence of Eqs.~(\ref{SOL1}) and~(\ref{SOL2}), which, applied to Eq.~(\ref{DEFdet}), gives
\begin{equation}\label{DEFdet1the}
\left.D\right\rvert_{F=0, T} 
\equiv D_{\rm th}=\left(1+ 2 \left<b^\dagger b\right>_{\rm th} \right)^2~.
\end{equation}
Therefore,
 \begin{equation}\label{DECERGOTHER1}
\left.\mathcal{E}_{\rm B}(\tau)\right\rvert_{F=0, T}
=\omega_0 \left.\left(\left<b^\dagger b\right>_{\rm th}-\frac{\sqrt{D}_{\rm th}-1}{2}\right)\right|_{t=\tau}=0~.
\end{equation}

To prove Eq.~(\ref{DECERGOCOE}), instead, we observe that  
 Eqs.~(\ref{eq:hoho-lang1co}) and~(\ref{eq:hoho-lang-in-condco})
 admit solutions for the second-order momenta which can be written as products of those obtained for the first-order momenta, i.e.
\begin{eqnarray}\label{fact}
\langle x y \rangle_{\rm co} &=& \langle x \rangle_{\rm co}  \langle y \rangle_{\rm co}~,
\end{eqnarray}
 for all  $x, y \in \{ a, b, a^\dag, b^\dag\}$.  Eq.~(\ref{fact}) implies that, during the time evolution, the state of the joint  system AB 
 is  described by a product state of the form
 \begin{eqnarray} 
\tilde{\rho}_{\rm AB}(t) = \ket{\alpha(t)}_{\rm A}\bra{\alpha(t)} \otimes \ket{\beta(t)}_{\rm B}\bra{\beta(t)}\;, \label{FACT1}
 \end{eqnarray}
 with $\ket{\alpha(t)}_{\rm A}$ and $\ket{\beta(t)}_{\rm B}$ coherent states of amplitudes $\alpha(t) \equiv \langle a\rangle_{\rm co}$ and 
 $\beta(t)\equiv \langle b \rangle_{\rm co}$, respectively. This result could have been anticipated by noting that 
 the ME for our model at $T=0$ induces an evolution of the input vacuum state through the combined action of a
  purely lossy channel and a displacement operator~\cite{GAUSSIANCV}. 
Using Eq.~(\ref{fact}) together with (\ref{DEC1rew}) and Eqs.~(\ref{SOL1})-(\ref{SOL2}),
it follows that the function (\ref{DEFdet}) for generic values of $F$ and $N_b(T)$ can be expressed as
\begin{equation}\label{DEFdet1}
D=\left(1+ 2 \left<b^\dagger b\right>_{\rm th}\right)^2~,
\end{equation}
with no dependence from contributions associated with the coherent driving. Accordingly,
replacing (\ref{DEFdet1})  into (\ref{eq:ergo-hoho}), we conclude that
\begin{widetext}
 \begin{equation}\label{DECERGOCOHE}
\mathcal{E}_{\rm B}(\tau)\Big\rvert_{F, T}
=\omega_0 \left.\left(\left<b^\dagger b \right>_{\rm th}+\left<b^\dagger b \right>_{\rm co}-\frac{\sqrt{D}-1}{2}\right)\right|_{t=\tau}=
\omega_0  \left<b^\dagger b \right>_{\rm co} = \left.{E}_{\rm B}(\tau)\right\rvert_{F, T=0}~,
\end{equation}
\end{widetext}
proving Eq.~(\ref{DECERGOCOE}).
 
 It is worth stressing that all the identities derived so far also hold for the local energy $E_{\rm A}(\tau)\equiv {\rm tr}[{H}_{\rm A} \rho_{\rm A}(\tau)]$ and the ergotropy  $\mathcal{E}_{\rm A}(\tau)\equiv E_{\rm A}(\tau)-E^{(p)}_{\rm A}(\tau)$ of the ancillary system $A$, i.e. explicitly
\begin{eqnarray} 
&&\left.E_{\rm A}(\tau)\right\rvert_{F, T} = \left.E_{\rm A}(\tau)\right\rvert_{F=0, T} + \left.E_{\rm A}(\tau)\right\rvert_{F, T=0}~, 
\nonumber \\
&&\left.\mathcal{E}_{\rm A}(\tau)\right\rvert_{F, T} = \left.{E}_{\rm A}(\tau)\right\rvert_{F, T=0}~, \nonumber \\ 
&&\left. \mathcal{E}_{\rm A}(\tau)\right\rvert_{F=0, T}=0~, \label{DECEA}
\end{eqnarray}
the first being just a trivial consequence of Eq.~(\ref{DEC1rew}) for $x=a^\dag a$, while the last two following from arguments similar to those we have adopted in deriving Eqs.~(\ref{DECERGOTHER}) and~(\ref{DECERGOCOHE}).

\section{Solving the ME for the two-qubit model}
\label{appendix-QubQub-me}

In order to solve Eq.~(\ref{eq:at-at-lindblad-me})
we expand all the operators appearing in it by utilizing a 
 global basis set for the two-qubit system $\{\ket{\ket{i}} \}_{i\in \{1,...,4\}}$.
We choose
 \begin{eqnarray}\label{eq:qubqub-open-collective-basis}
 \ket{\ket{1}}&=&\ket{1}_{\rm A }\ket{1}_{\rm  B}~, \quad \ket{\ket{2}}=\ket{1}_{\rm A }\ket{0}_{\rm  B}~,\\\nonumber
 \ket{\ket{3}}&=&\ket{0}_{\rm A }\ket{1}_{\rm  B}~,\quad \ket{\ket{4}}=\ket{0}_{\rm A }\ket{0}_{\rm  B}~,
\end{eqnarray}
where $\ket{1}_{\rm  A(B)}$ and $\ket{0}_{\rm  A (B)}$ are the eigenvectors of the $\sigma^z_{\rm A (B)}$ operators with eigenvalues $\pm 1$. Accordingly, we write 
$\tilde{\rho}_{\rm A B}(t) =\sum_{i,j=1}^4 r_{ij}(t) \ket{\ket{i}}\bra{\bra{j}}$, or, in matrix form,
\begin{eqnarray}\label{eq:r_ij}
\tilde{\rho}_{\rm A B}(t)\equiv
\begin{pmatrix}
r_{11}(t) &r_{12}(t) &r_{13}(t) &r_{14}(t) \\
r_{21}(t) &r_{22}(t) &r_{23}(t) &r_{24}(t) \\
r_{31}(t) &r_{32}(t) &r_{33}(t) &r_{34}(t) \\
r_{41}(t) &r_{42}(t) &r_{43}(t) &r_{44}(t) \\ 
\end{pmatrix}~,
\end{eqnarray}
$r_{ij}(t)$  being expansion coefficients. In this representation, the ladder operators $a,a^\dagger$ of the subsystem A can instead be written as 
\begin{eqnarray}
\sigma^-_{\rm{A}}\equiv  
\left(
\begin{array}{c|c}
\mathbf{0} & \mathbf{0} \\
\hline
\mathbb{1} & \mathbf{0}
\end{array}
\right)\\
\sigma^+_{\rm{A}}\equiv  
\left(
\begin{array}{c|c}
\mathbf{0} & \mathbb{1} \\
\hline
\mathbf{0} & \mathbf{0}
\end{array}
\right)
\end{eqnarray}
where $\mathbf{0}$ is the $2\times 2$ matrix with all null entries.
Finally, the system Hamiltonian (up to an irrelevant additive constant) is represented by 
\begin{eqnarray}
{H}(t)\equiv 
\begin{pmatrix}
 \omega_0 & 0 & F e^{- i \omega_0 t}  & 0 \\
 0 & 0 & g & F e^{- i \omega_0 t}\\
 F^* e^{ i \omega_0 t}& g  & 0  & 0 \\
 0 & F^* e^{ i \omega_0 t} & 0 & -\omega_0\ 
\end{pmatrix}~.
\end{eqnarray}
With these choices, Eq.~(\ref{eq:at-at-lindblad-me}) translates into a first-order system of ordinary differential equations in the sixteen unknown functions $r_{\rm i j}(t)$, which has to be solved under the initial conditions (\ref{eq:in-cond-qubqub}) corresponding to 
$r_{ij}(0)=1$ for $i=j=4$ and zero otherwise.

Explicit expressions for the local energies of A and B can be obtained once the operators $\sigma_z^{(\rm A)}$ and  $\sigma_z^{(\rm B)}$ are represented in the basis (\ref{eq:qubqub-open-collective-basis}). It turns out that they take the following forms
\begin{equation}
E_{\rm A}(\tau) = \frac{\omega_0}{2}\left[r_{11}(\tau)+r_{22}(\tau)-r_{33}(\tau)-r_{44}(\tau)+1 \right]
\end{equation}
and
\begin{equation}
E_{\rm B}(\tau) = \frac{\omega_0}{2}\left[r_{11}(\tau)-r_{22}(\tau)+r_{33}(\tau)-r_{44}(\tau)+1\right]~.
\end{equation}
Finally, the ergotropy of B reads as following
\begin{eqnarray}
\mathcal{E}_{\rm B}(\tau)&=&\frac{\omega_0}{2}\times\nonumber\\
&\times&\{\sqrt{4 |r_{12}+r_{34}|^2+[2 (r_{11}+r_{33})-1]^2}\nonumber\\
&+&2 (r_{11}+r_{33})-1\}~.
\end{eqnarray}
\end{document}